%% file: main_r2.tex
\documentclass[preprint,10pt]{elsarticle}                                                                                                                                               
\usepackage[displaymath]{preview}
\usepackage{graphicx}
\usepackage{cancel}
\usepackage{latexsym}
\usepackage{subfigure}
\usepackage{epsfig}
\usepackage{amsmath}
\usepackage{rotating}
\usepackage{enumerate}
\usepackage{multirow}
\usepackage{amssymb}
\usepackage{tikz}
\usetikzlibrary{calc}
\usepackage[mathscr]{eucal}
\usepackage{amsthm}
\usepackage{units}
\usepackage{pgfplots}
\usepackage{color} %\usepackage[small]{caption}                                                                                                                                         
\usepackage[hypertexnames=false,colorlinks=true,linkcolor=blue,citecolor=blue,                                                                                                          
bookmarksopen=false,bookmarks=false,                                                                                                                                                    
pdfstartview=XYZ,pdffitwindow=true,pdfcenterwindow=true]{hyperref}
\usepackage[a4paper,text={6.0in,9.0in},centering,                                                                                                                                       
includefoot,foot=0.6in]{geometry} %\usepackage[printfigures]{figcaps}                                                                                                                   
\usepackage{pdfsync} 

\newcommand{\x}{\mathbf{x}}
\newcommand{\y}{\mathbf{y}}
\newcommand{\z}{\mathbf{z}}
\allowdisplaybreaks[1]

\usepackage[english]{babel}
\usepackage{fontenc}
\usepackage{amsfonts}
\usepackage{bm}
\usepackage{amsthm}
\usepackage{todonotes}

\begin{document}
\title{A fast and robust computational method for the ionization cross sections of the driven Schr\"odinger equation using an $\mathcal{O}(N)$ multigrid-based scheme}
                                                                                                                            
\author[ua]{S. Cools}  \ead{siegfried.cools@uantwerpen.be}
\author[ua]{W. Vanroose} \ead{wim.vanroose@uantwerpen.be}
\address[ua]{Applied Mathematics Research Group, Department of Mathematics and Computer Science, University of Antwerp, Middelheimlaan 1, 2020 Antwerp, Belgium} 

\date{\today}

\begin{abstract} 
This paper improves the convergence and robustness of a multigrid-based
solver for the cross sections of the driven Schr\"odinger equation.  
Adding an Coupled Channel Correction Step (CCCS) after each multigrid 
(MG) V-cycle efficiently removes the errors that remain after the V-cycle sweep.
The combined iterative solution scheme (MG-CCCS) is shown to feature 
significantly improved convergence rates over the classical MG method 
at energies where bound states dominate the solution, resulting in a 
fast and scalable solution method for the complex-valued Schr\"odinger 
break-up problem for any energy regime. 
The proposed solver displays optimal scaling; 
a solution is found in a time that is linear in the number of unknowns.
The method is validated on a 2D Temkin-Poet model problem, and
convergence results both as a solver and preconditioner are provided
to support the $\mathcal{O}(N)$ scalability of the method. 
This paper extends the applicability of the complex contour approach 
for far field map computation [\emph{S.~Cools, B.~Reps, W.~Vanroose, 
An Efficient Multigrid Calculation of the Far Field Map for Helmholtz 
and Schr\"odinger Equations, SIAM J.~Sci.~Comp.~36(3) B367--B395, 2014}].

%% This paper extends the complex contour approach for far field map
%% computation [\emph{S.~Cools, B.~Reps, W.~Vanroose, An Efficient
%%     Multigrid Calculation of the Far Field Map for Helmholtz and
%%     Schr\"odinger Equations, SIAM J.~Sci.~Comp.~36(3) B367--B395
%%     (2014)}] that allows for a highly efficient multigrid-based
%% calculation of the far field map for Helmholtz and Schr\"odinger type
%% scattering problems. With this work we aim at further improving the
%% proposed iterative multigrid solver by adding an advanced Coupled
%% Channel Correction Step (CCCS) after each multigrid (MG) V-cycle,
%% which is designed specifically to account for the presence of
%% localized bound states in the solution.  The combined iterative
%% solution scheme (denoted by MG-CCCS) is shown to feature significantly
%% improved convergence rates over the classical MG method at energies
%% where bound states dominate the solution, resulting in a fast and
%% scalable solution method for the Schr\"odinger break-up problem for
%% any energy regime. Numerical results show that this leads to a robust
%% method for the computation of the ionization cross sections for
%% electron-impact models.  The proposed MG-CCCS method is validated on a
%% 2D Temkin-Poet model problem, and convergence results using the
%% MG-CCSS scheme both as a solver and preconditioner are provided to
%% support the $\mathcal{O}(N)$ scalability of the method.
\end{abstract}

\maketitle

%\textbf{Keywords:} Schr\"odinger equation, cross sections, complex contour, multigrid, coupled channel correction scheme, Temkin-Poet

%%%%%%%%%  SECTION  %%%%%%%%%%%%%%%%%%%%%%%%%%%%%%%%%%%%%%%%%%%%%%%%

\section{Introduction} \label{sec:intro}

Over the past decades significant work has been performed on the
computational simulation of breakup problems in atomic and molecular
physics.  The cross sections of electron-impact ionization
problems were calculated for atoms and molecules in
\cite{mccurdy1997approach,rescigno1999collisional,baertschy2001electron,mccurdy2004solving,vanroose2006double}.
However, the classical algorithms for the computation of the
ionization cross sections for a general Schr\"odinger type scattering
problem are notably very demanding in terms of computational
resources, and hence require supercomputing infrastructure. This
computational cost primarily originates from the computation of
the high-dimensional scattering solutions governed by Schr\"odinger's
equation, which are notoriously hard to obtain using either classical
direct methods or iterative solvers.  Hence, numerical simulations of
the far field behaviour for molecular break-up problems are generally
computationally challenging.  The aim of this work is to design an
efficient and scalable method for the computation of the cross
sections of the $d$-dimensional driven Schr\"odinger equation.
Moreover, we aim at computing the single differential cross section
(SDCS) and total cross section (TCS) for Schr\"odinger type scattering
problems using a multigrid-based method that scales linearly in
the number of unknowns.
	
Multigrid \cite{brandt1977multi} is an efficient iterative solution
method for the large and sparse linear systems that appear in the
numerical solution of Laplace-type partial differential equations
(PDEs). The method relies upon a hierarchy of discretizations of the
problem to accelerate convergence to the solution at the finest
discretization level.  The key concept of the multigrid method is the
synergy between the \emph{smoother}, i.e.~a basic iterative update
step such as weighted Jacobi or Gauss-Seidel that is applied at each
level, and a coarse grid correction scheme that solves the error
equation on a coarser discretization level.  Consequently, the
interpolated error is used to correct the fine-level guess to the
solution \cite{brandt1977multi,briggs2000multigrid}.  From a spectral
analysis point of view, the smoother removes the oscillatory parts
from the error, resulting in a remaining error that is smooth enough
to be represented at the coarser levels of the discretization
hierarchy.  Applying this recursively leads to an
$\mathcal{O}(N)$ scheme for the solution of Laplace-type PDEs, where
$N$ is the number of unknowns at the finest level.  The multigrid
method is therefore commonly used to solve high-dimensional Poisson
problems, where it is particularly efficient due to its
scalability. Additionally, multigrid is in practice often used as a
preconditioner to accelerate the convergence of Krylov subspace
iterations such as Conjugate Gradients (CG) or GMRES
\cite{van2003iterative}.  Multigrid-preconditioned Krylov methods have
been successfully applied as solvers for e.g.~fluid flow models, see
\cite{elman2011fast}.  Moreover, it has been shown in the literature
that the use of multigrid as a Krylov method preconditioner results in
an efficient and scalable solution method for Laplace-type problems,
featuring a convergence rate which is independent of the number of
unknowns \cite{trottenberg2001multigrid}.
	
Although multigrid is customarily applied as both a solver and
preconditioner for a wide range of Laplace-type PDEs, it is well-known
that the classical multigrid method (like any iterative solver)
utterly breaks down when applied to Helmholtz or Schr\"odinger
problems. The reason for this convergence failure is best understood
by analyzing the spectral properties of the corresponding operator.
Due to the intrinsic indefiniteness of the discretized Helmholtz
operator, the occurrence of near-zero coarse level eigenvalues
destroys the stability of both smoother and two-grid correction
scheme, leading to a possibly highly instable multigrid scheme
\cite{elman2002multigrid,ernst2010difficult}.  In the past decades
many efforts have been performed to overcome these stability issues of
multigrid for Helmholtz equations. It has been proposed in
\cite{erlangga2006novel} and \cite{made2001incomplete} that multigrid and ILU can be
successfully used as a solver for the complex shifted (i.e.~damped)
Helmholtz equation. This perturbed problem is known as \emph{complex
  shifted Laplacian} (CSL), and is commonly used as a preconditioner
for Helmholtz problems \cite{bollhoefer2009algebraic}.  Alternatively,
the underlying numerical grid can be rotated into the complex domain
to yield an equivalent damped Helmholtz problem, which is known as
\emph{complex stretched grid} (CSG) \cite{reps2010indefinite}. This
technique shows significant resemblances to \emph{Exterior Complex
  Scaling} (ECS) \cite{simon1979definition}.

In this paper the properties of the multigrid method on damped
Helmholtz and Schr\"odinger equations are exploited to construct an
efficient computational method for the far field map (ionization cross
sections) of Helmholtz (Schr\"odinger) type scattering problems. The
computation of the far field map can typically be seen as a two-step
process. The first step consists of solving the underlying Helmholtz
or driven Schr\"odinger equation on a bounded numerical domain
with absorbing boundary conditions, whereas in the second
step an integral over the scattering solution on this domain has to be
calculated \cite{kirsch1996introduction, colton1998inverse}. The main
computational bottleneck is generally the first step, since this
requires the solution of a large scale scattering problem that is
typically hard to obtain using classical iterative methods.  However,
in \cite{cools2014efficient} it was proposed that the classical far
field integral over the real-valued $d$-dimensional bounded domain can
alternatively be replaced by an integral over a complex-valued
manifold. Furthermore, the proposed technique was validated
\cite{cools2014efficient} by showing that the complex contour approach
yields an identical far field map. Moreover, after the deformation of
the integral, the first step in the calculation now consists of
solving a damped Helmholtz or Schr\"odinger equation, which can be
very efficiently achieved using e.g.~multigrid methods. The main
computational bottleneck in the far field map computation is hence
overcome, since the numerical solution on the complex domain can
effectively be obtained using a scalable iterative method. However,
it was reported in \cite{cools2014efficient} that the proposed method 
fails for a range of negative energies where single ionization is present.
	
In this work we extend the complex contour approach introduced in
\cite{cools2014efficient} to compute the ionization cross sections of
2D Schr\"odinger type scattering problems, such that it works for all 
energies, including those energy regimes where single ionization waves 
dominate the solution. The multigrid method used
for the solution of the complex-valued driven Schr\"odinger problem
(step 1) is extended by the addition of a so-called \emph{Coupled
  Channel Correction Scheme} (CCCS)
\cite{heller1973comment,mccarthy1983momentum}, which ensures the
efficiency of the proposed solver for all possible energy levels by
specifically eliminating the bound states, i.e.~low-dimensional
localized waves travelling along the domain edges.  The Coupled
Channel Correction Scheme is derived analytically, and we expound on
its numerical properties through an eigenvalue analysis.  Numerical
results in 2D are provided to illustrate that the combination of
multigrid and the Coupled Channel Correction Scheme (MG-CCCS)
effectively leads to a fast, scalable and robust iterative scheme for
solving the driven Schr\"odinger equation for any energy regime.
	
%% We additionally comment on the choice of the complex rotation angle in
%% this work. Although from the analytical point of view there are no
%% fundamental restrictions on the magnitude of the complex rotation, and
%% the multigrid solver clearly benefits from a heavy damping of the
%% problem, we show that choosing the rotation angle too large may result
%% in numerical instabilities when calculating the integral in the second
%% step of the far field map computation. Indeed, the choice of the rotation angle is
%% shown to be restricted primarily by the numerical implementation, and is thus in
%% practice subject to a trade-off between multigrid performance for computing the
%% scattered wave solution (step 1) and computational stability of the
%% far field integral (step 2).

%%  Moved down:
%% 	
%% The Temkin-Poet (TP) model will be used throughout this work as an
%% approximation to an impact ionization problem, where only s-waves are
%% used to describe the scattered electrons. The model has a long
%% history as a benchmark problem for Schr\"odinger
%% scattering problems, see e.g.\ \cite{rescigno1999collisional}, where a
%% two-dimensional TP model was used to describe the impact ionization of
%% molecular hydrogen.  Similarly, the impact ionization of helium can be
%% modelled by a 3D Temkin-Poet model \cite{vanroose2006double}. 
%% The TP model problem serves as the primary test case
%% for numerical validation of the proposed MG-CCCS method in this paper.

The outline of the article is as follows. Section \ref{sec:schro} introduces
the general notations and basic concepts used throughout this
work. In Section \ref{sec:compl} we briefly revisit the complex contour method for
the computation of the far field map, which was initially proposed in
\cite{cools2014efficient}, and we comment on the spectral
properties of the complex rotated Schr\"odinger equation.  This
discussion motivates the need for the so-called Coupled Channel
Correction Scheme (CCCS) introduced in Section \ref{sec:multi}, which is used to
compensate for the deterioration of multigrid convergence in certain
energy regimes where bound states dominate the solution. Section \ref{sec:imple}
expounds on the implementation of the Coupled Channel scheme and the
influence of the complex rotation angle on numerical accuracy.  Taking
into account these remarks, the single differential and total cross
sections of a 2D Temkin-Poet model problem are computed using the
MG-CCCS method.  We illustrate the efficiency of the combined MG-CCCS
method both as a solver and a preconditioner in Section \ref{sec:numer}.
Furthermore, the robustness of the MG-CCCS preconditioned Krylov
solvers with respect to the energy of the system is demonstrated in this section. Section \ref{sec:concl} concludes
this work alongside a discussion on the topic.

%%%%%%%%%  SECTION  %%%%%%%%%%%%%%%%%%%%%%%%%%%%%%%%%%%%%%%%%%%%%%%%

\section{The Schr\"odinger equation and the far field map} \label{sec:schro}

The driven time-independent Schr\"odinger equation describing a system 
consisting of one positively charged nucleus located at the origin and 
two negatively charged electrons with spatial coordinates $\mathbf{r}_1$
and $\mathbf{r}_2 \in \mathbb{R}^3$ is given by
\begin{equation} \label{eq:schrodinger}
 (\mathcal{H}-E) \, \psi(\mathbf{r}_1, \mathbf{r}_2) = \xi(\mathbf{r}_1,\mathbf{r}_2).
\end{equation}
The operator $\mathcal{H}$ denotes 
the Hamiltonian
\begin{equation*}%\label{eq:schrodinger_hamiltonian}
  \mathcal{H} =  - \frac{1}{2m} \triangle_{\mathbf{r}_1}  - \frac{1}{2m} \triangle_{\mathbf{r}_2}  + V_1(\mathbf{r}_1) + V_2(\mathbf{r}_2) + V_{12}(\mathbf{r}_1,\mathbf{r}_2),
\end{equation*}
where the mass $m > 0$ scales the Laplacians.  The right hand side
$\xi(\mathbf{r}_1,\mathbf{r}_2)$ in \eqref{eq:schrodinger} can model
either an incoming electron that impacts the system
\cite{rescigno1999collisional}, or alternately, represents the dipole
operator working on a ground state if the model is used to compute
photoionization \cite{vanroose2005complete}. For these break-up
problems the solution $\psi(\mathbf{r}_1,\mathbf{r}_2)$ is an outgoing
wave in any direction and leads to a six-dimensional problem on an
unbounded domain.  Note that in the context of simulating electron
impact ionization the two electrons, namely the electron particle in the
target and the impinging electron, are indistinguishable and need to
be described on equal footing. This naturally leads to the 6D
Schr\"odinger equation \eqref{eq:schrodinger}. The potential is
typically split into a sum of one-body potentials $V_1$ and $V_2$, and
a two-body potential $V_{12}$. Here the one-body potentials model the
attraction of each electron by the nuclear charge, whereas the
two-body potential simulates the electron-electron repulsion.  More
generally, for a system containing $d$ particles, the potential
$V(\mathbf{r}_1,\ldots,\mathbf{r}_d)$ can be written as
\begin{equation} \label{eq:potential_split}
  V(\mathbf{r}_1,\ldots,\mathbf{r}_d) = \sum_{i=1}^d V_i(\mathbf{r}_i) + \sum^d_{i<j} V_{ij}(\mathbf{r}_i,\mathbf{r}_j) = \bar{V}_1(\mathbf{r}_1,\ldots,\mathbf{r}_d) + \bar{V}_2(\mathbf{r}_1,\ldots,\mathbf{r}_d).
\end{equation}
We use $\bar{V}_1$ to denote the sum of all one body potentials and
$\bar{V}_2$ for the sum of the two-body potentials throughout the following sections.

%---------------- SUBSECTION --------------------------------

\subsection{The partial wave expansion}

At large distances from the object, the solution of
\eqref{eq:schrodinger} behaves like a spherical wave emerging from the
center of mass of the system.  Equation \eqref{eq:schrodinger} is
therefore commonly rewritten in spherical coordinates
\cite{baertschy2001electron,vanroose2006double}, where the Laplacian
operator splits into a radial operator and an angular differential
operator \cite{arfken2005mathematical}. The coordinates are written as
$\mathbf{r}_1 = (\rho_1, \Omega_1)$ and $\mathbf{r}_2 = (\rho_2,
\Omega_2)$, where $\Omega_i$ denotes the solid angle in the
three-dimensional space.  The solution is represented by the series
\begin{equation}\label{eq:proposal}
\psi(\mathbf{r}_1,\mathbf{r}_2) = \sum_{l_1=0}^{\infty}\, \sum_{m_1=-l_1}^{l_1}\,\sum_{l_2=0}^\infty\, \sum_{m_2=-l_2}^{l_2}  \psi_{l_1m_1,l_2m_2}(\rho_1,\rho_2) Y_{l_1m_1}(\Omega_1) Y_{l_2m_2}(\Omega_2), 
\end{equation}
where $Y_{l_im_i}(\Omega_i)$ are the spherical harmonics, i.e.\ the eigenfunctions of the angular 
differential operator of the Laplacian in spherical coordinates \cite{arfken2005mathematical}.
The above decomposition is commonly referred to as the partial wave expansion.
After substitution of \eqref{eq:proposal} into the Schr\"odinger equation \eqref{eq:schrodinger} and an integration over $\Omega_1$ and $\Omega_2$
one obtains an equation for the radial components $\psi_{l_1\,m_1,l_2\,m_2}(\rho_1,\rho_2)$ for all 
$l_1 \ge 0,l_2 \ge 0 ,|m_1| \le l_1 ,|m_2| \le l_2$, that is coupled to all other partial waves, i.e.
\begin{equation}\label{eq:coupled}
\begin{aligned}
 &\left[-\frac{1}{2m}  \triangle_{l_1,l_2} -E\right]  \psi_{l_1 m_1,l_2 m_2}(\rho_1,\rho_2) \\
&+ \sum_{l^{\prime}_1=0}^{\infty} \sum_{m^{\prime}_1=-l^{\prime}_1}^{l^{\prime}_1} \sum_{l_2^\prime = 0}^\infty \sum_{m^\prime_2=-l_2^\prime}^{l_2^\prime} V_{l_1\,m_1\,l_2\,m_2 ; l_1^\prime\,m_1^\prime,l_2^\prime\, m_2^\prime}(\rho_1,\rho_2)\,\, \psi_{l_1^\prime m_1^\prime,l_2^\prime m_2^\prime}(\rho_1,\rho_2)  = \xi_{l_1m_1,l_2m_2}.
\end{aligned}
\end{equation}
Note that the system decouples when the potentials $V_1$, $V_2$ and $V_{12}$ are spherically symmetric.
The differential operators only appear on the diagonal blocks of the resulting coupled system.
These diagonal blocks take the form of a $2$-dimensional driven Schr\"odinger equation 
\begin{equation} \label{eq:schrod}
  \left(-\frac{1}{2} \Delta  + V(\mathbf{x}) - E\right) u(\mathbf{x}) = f(\mathbf{x}), \quad \text{for} ~ \textbf{x} = (x,y) \in \mathbb{R}^2, 
\end{equation}
for a system with unit mass, where $\Delta$ is the 
$2$-dimensional Laplacian, $V(\mathbf{x})$ is a
scalar potential, $E$ is the total energy of the system, $u$ is the outgoing
partial wave function and $f$ is the right-hand side, representing the
effect of an incoming wave (electron scattering) or the influence of an optical perturbation
such as interaction with light (photoionization). The operator 
$H = -\frac{1}{2} \Delta+ V(\mathbf{x})$ is the Hamiltonian. The potential 
$V(\mathbf{x})$ includes angular potentials like $l_i(l_i +1)/x^2$ 
$(i = 1,2)$, related to the partial wave angular momentum.
Note that the above formulation directly extends to the case where three or more
particles are considered. Equation \eqref{eq:schrod} then generalizes to  
\begin{equation} \label{eq:schrod2}
  \left(-\frac{1}{2} \Delta^d  + V(\mathbf{x}) - E\right) u(\mathbf{x}) = f(\mathbf{x}), \quad \text{for} ~ \textbf{x} \in \mathbb{R}^d, 
\end{equation}
where $d = 2,3,\ldots$ corresponds to the number of electrons in the system,
and hence the dimension of the radial partial wave equation.
The operator $\Delta^d$ is the related $d$-dimensional partial wave Laplacian.

%---------------- SUBSECTION --------------------------------

\subsection{The 2D Schr\"odinger equation: notation and basic properties}

As suggested above, for the 6D scattering problem \eqref{eq:schrodinger} the 
diagonal blocks of the coupled system \eqref{eq:coupled} take the form of a 
two-dimensional Schr\"odinger equation
\begin{equation}\label{eq:partialwave2d}
\left(-\frac{1}{2} \Delta  + V_1(x) + V_2(y) + V_{12}(x,y) -E\right) u(x,y) = f(x,y),   \quad x,y \ge 0,
\end{equation}
with boundary conditions
\begin{equation*}%\label{eq:partialwave2dBC}
\begin{cases} 
u(x,y)=0 \quad & \text{for} \quad x = 0 \quad \text{or} \quad y = 0, \\
\text{outgoing} \quad & \text{for} \quad x \rightarrow \infty \quad \text{or} \quad y \rightarrow \infty,
\end{cases}
\end{equation*}
where $V_1(x)$ and $V_2(y)$ are the one-body potentials and $V_{12}(x,y)$ 
is a two-body potential. Since the arguments $x$ and $y$ are in fact radial 
coordinates in the partial wave expansion, homogeneous Dirichlet boundary 
conditions are imposed at the $x=0$ and $y=0$ boundaries. The potentials 
$V_1$, $V_2$ and $V_{12}$ are generally analytical functions that decay as 
the radial coordinates $x$ and $y$ become large. 
%In the case of the 9D 
%scattering problem, the model problem derived through a partial wave expansion
%gives rise to a 3D Schr\"odinger problem
%\begin{align}\label{eq:partialwave3d}
%  \Big(-\frac{1}{2} \Delta  &+ V_1(x) +  V_2(y) + V_3(z) \notag \\
%  &   + V_{12}(x,y) + V_{23}(y,z)+ V_{13}(z,x)-E \, \Big) \, u(x,y,z) = f(x,y,z),  \quad x,y,z \ge 0,
%\end{align}
%with boundary conditions
%\begin{equation}\label{eq:partialwave3dBC}
%\begin{cases} 
%u(x,y,z)=0 \quad &  \text{for} \quad x = 0 \quad \text{or} \quad y = 0 \quad \text{or} \quad z = 0,\\
%\text{outgoing} \quad & \text{for} \quad x \rightarrow \infty \quad \text{or} \quad y \rightarrow \infty \quad \text{or} \quad z \rightarrow \infty,
%\end{cases}
%\end{equation}

%---------------- SUBSECTION --------------------------------

\subsection{Relation to the Helmholtz equation}

Depending on the total energy of the system $E \in \mathbb{R}$, the
above Schr\"odinger system allows for scattering solutions, in which
case equation \eqref{eq:schrod2} can be reformulated as a $d$-dimensional Helmholtz
equation of the form
\begin{equation} \label{eq:helmholtz_ddim}
\left(-\Delta-k^2(\mathbf{x})\right) \, u(\mathbf{x}) = g(\mathbf{x}), \quad \text{for} ~ \textbf{x} \in \mathbb{R}^d, 
\end{equation}
where the spatially dependent wavenumber $k(\mathbf{x})$ is defined by
$k^2(\mathbf{x}) := 2(E-V(\mathbf{x}))$ and the right-hand side is
$g(\x) = 2f(\x)$.  The experimental observations from this type of
quantum mechanical systems are typically far field maps of the
solution \cite{vanroose2006double}.  In many quantum
mechanical systems the potential $V(\mathbf{x})$ is an analytical
function, which suggests analyticity of the spatially dependent
wavenumber $k(\mathbf{x})$ in the above Helmholtz equation.
Additionally, the potential $V(\x)$ is often decaying in function 
of growing $\x$. These observations will prove particularly valuable in 
the context of this paper.

%---------------- SUBSECTION --------------------------------

\subsection{The single and double ionization regimes}

\begin{figure}
\begin{center}
\begin{tabular}{cc}
 \includegraphics[width=0.45\textwidth]{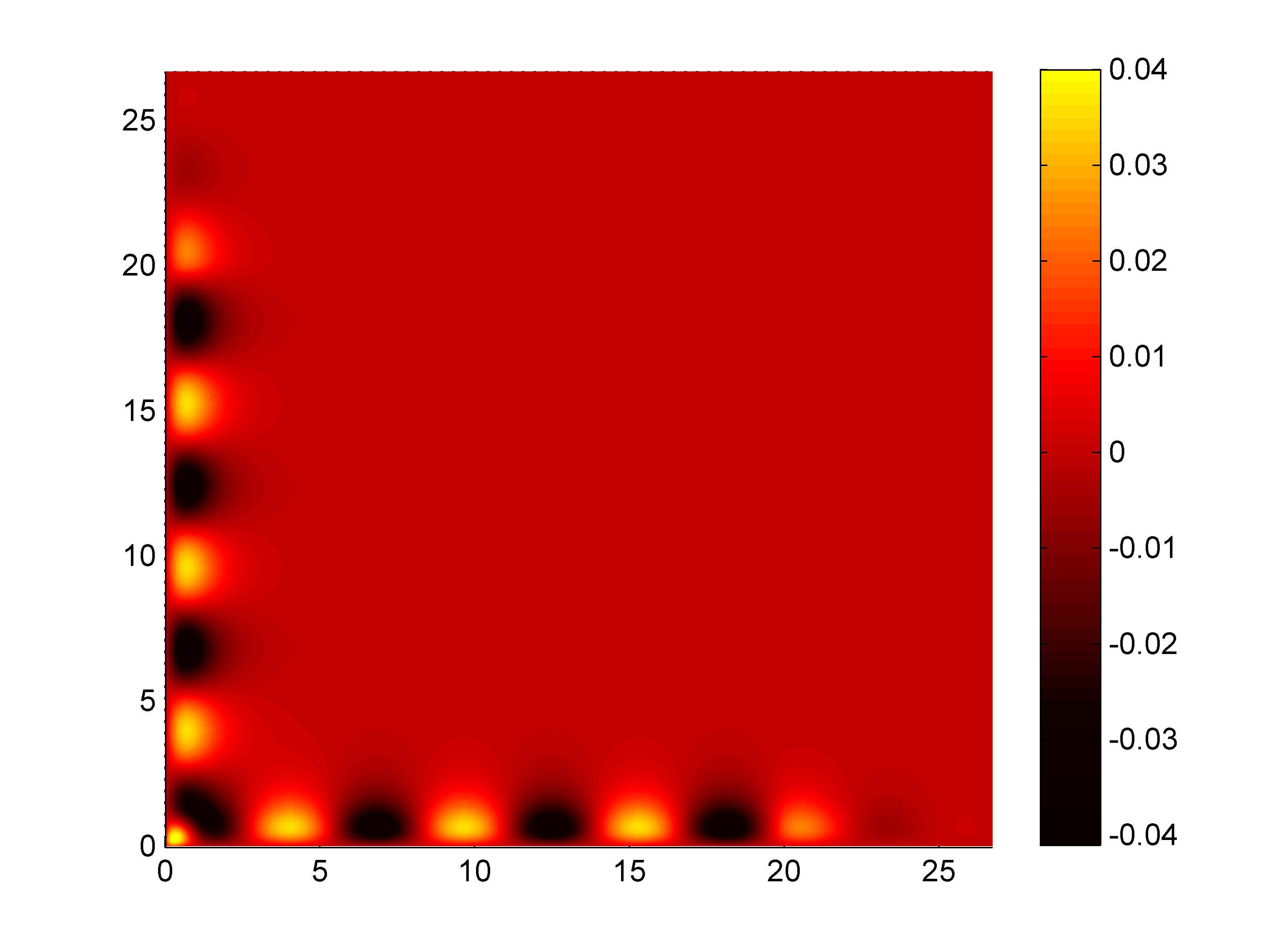} & 
 \includegraphics[width=0.45\textwidth]{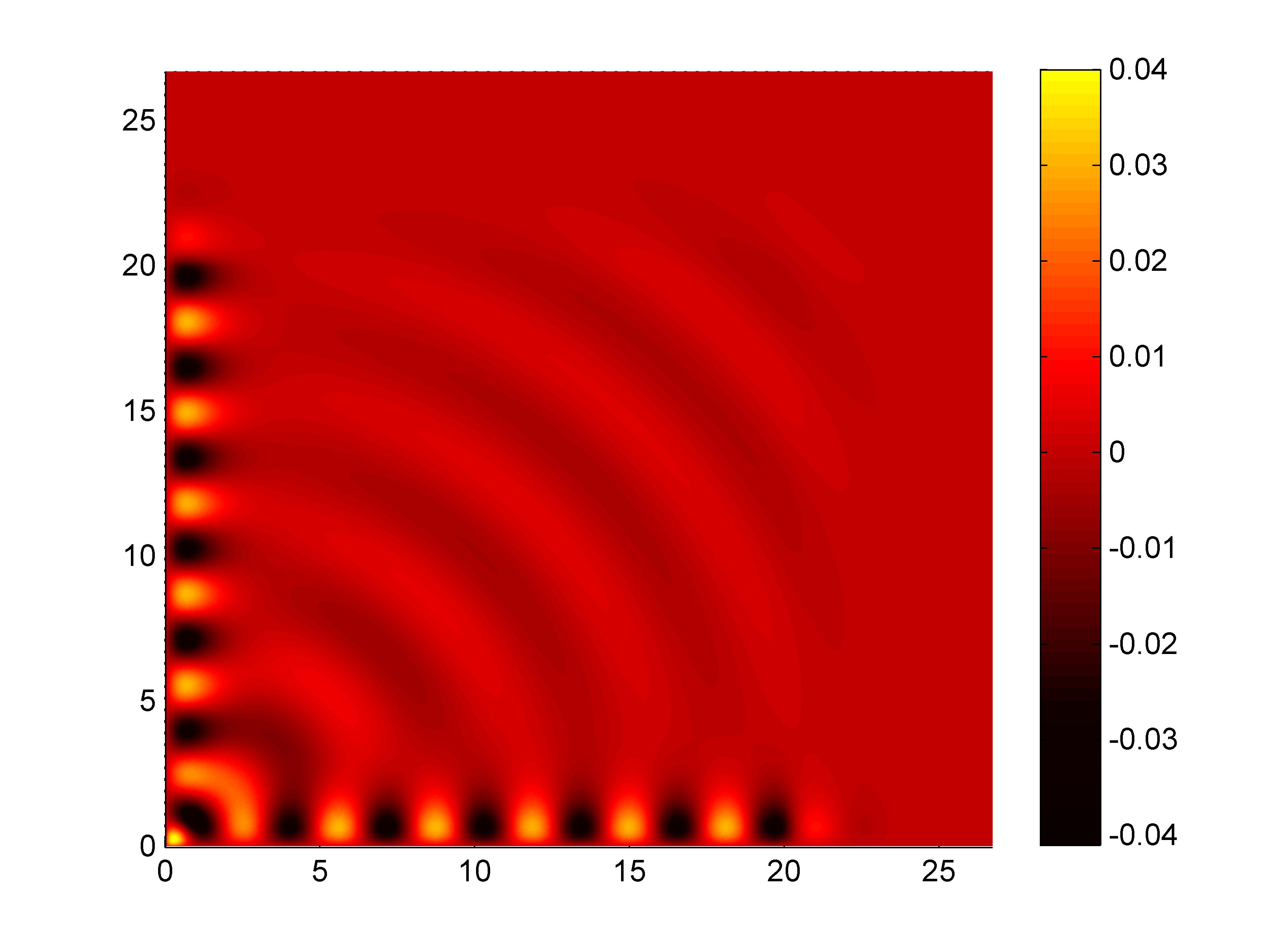} \\ 
 {\footnotesize (a) Single ionization ~~~~~~} & {\scriptsize (b) Double ionization ~~~~~~} \vspace{-0.3cm} 
\end{tabular}
\end{center}
\caption{Scattered wave solutions $u(x,y)$ to model problem \eqref{eq:partialwave2d}. 
	Model specifications: finite difference discretization on the domain $\Omega = [0,20]^2$
	using $n_x \times n_y = 400 \times 400$ grid points (+ 150 points per dimension in the ECS absorbing layer).
	Exponential potentials $V_1(x) = -4.5\exp(-x^2)$, $V_2(y) = -4.5\exp(-y^2)$ 
	and $V_{12}(x,y) = 2\exp(-(x+y)^2)$ and right-hand side $f(x,y) = \exp(-3(x+y)^2)$.
	(a): Solution for energy $E=-0.4$ for which only single ionization occurs. 
	Single ionization waves are localized bound states travelling along the edges of the domain.
  (b): Solution for energy $E=1$ for which both single and double ionization occur,
  resulting in the existence of waveforms along the edges as well as in the middle of the domain.}
\label{fig:single_and_double_sol}
\end{figure}

We now briefly expound on the physical interpretation of the 2D
Schr\"odinger system \eqref{eq:partialwave2d} in which $V_1$ and $V_2$
are assumed to be identical one-body potentials, which describe the
attraction (repulsion) to the central nucleus, and $V_{12}$ is a
two-body potential, which describes the mutual attraction (repulsion)
between the two scattered particles.  Depending on the strength of the
one-body potentials $V_1$ and $V_2$ and the energy $E$, the problem
allows for so-called single ionization waves, which are localized
evanescent waves that propagate along the edges of the domain, see
Figure \ref{fig:single_and_double_sol}(a). If the attraction of $V_1$
is strong enough, there exists a one-dimensional eigenstate
$\phi_n(x)$ with a negative eigenvalue $\lambda_n < 0$, characterized
by the one-dimensional Helmholtz equation
\begin{equation}\label{eq:1d_eigenstate}
  \left(-\frac{1}{2}\frac{d^2}{dx^2}+V_1(x)\right)\phi_n(x) = \lambda_n \, \phi_n(x),  \quad \text{for} ~ x \ge 0.
\end{equation}
Note that $\phi_n(0)=0$ and $\phi_n(x \rightarrow \infty)=0$.  If the
two-body potential $V_{12}(x,y)$ is negligibly small, then there
automatically also exists a bound state of the 2D system, since each
state $\phi_n(x)\phi_n(y)$ is an eigenstate of the separable
Hamiltonian $(-1/2) \Delta + V_1(x) + V_2(y)$ with eigenvalue
$2\lambda_n$.  In the presence of a small but non-negligible two-body
potential, this eigenstate will be slightly perturbed, resulting in an
eigenstate $\phi_n(x,y)$ with eigenvalue $\mu_n$ that fits the 2D subsystem
\begin{equation}\label{eq:2d_eigenstate}
\left(-\frac{1}{2} \Delta  + V_1(x) +  V_2(y) + V_{12}(x,y)\right) \phi_n(x,y) = \mu_n\phi_n(x,y),   \quad x,y \ge 0.
\end{equation}

The corresponding eigenvalue is $\mu_n \approx 2\lambda_n < \lambda_n
< 0$. This ordering is typical for realistic atomic and molecular
systems \cite{balslev1971spectral}.  
%Similarly, the 3D system will
%have an eigenstate that looks approximately like
%$\phi_n(x,y)\phi_n(z)$, or any of its coordinate permutations. The
%fully 3D eigenstate $\phi_n(x,y,z)$ fits the equation
%\begin{align} \label{eq:3d_eigenstate}
%  \Big( -\frac{1}{2} \Delta  &+ V_1(x) +  V_2(y) + V_3(z) \notag \\
%  &   + V_{12}(x,y) + V_{23}(y,z)+ V_{13}(z,x) \Big) \, \phi_n(x,y,z) = \nu_n \, \phi_n(x,y,z),  \quad x,y,z \ge 0,
%\end{align}
%where $\nu_n \approx \mu_n + \lambda_n \approx 3 \lambda_n$.  
Assuming that the potentials are such that $\mu_n < \lambda_n < 0$,
there are three possible regimes of interest in equation
(\ref{eq:partialwave2d}), depending on the total energy $E$.  First,
for $E < \lambda_n$, the problem is easy to solve numerically, but no
interesting physical reactions occur, since there are no scattering 
states in the solution.
For energy levels $\lambda_n < E < 0$, single ionization
scattering occurs, i.e.~there are scattering solutions that are
localized along one of the two axes in the 2D domain, 
see Figure \ref{fig:single_and_double_sol}(a). 
These solutions take the form $\phi(y)\phi_n(x)$ as $y \rightarrow \infty$,
where $\phi_n(x)$ is the eigenstate of \eqref{eq:1d_eigenstate} and
$\phi(y)$ is a scattering solution satisfying outgoing wave boundary
conditions, and vice versa for the permuted coordinates.
For energies $E > 0$, both single and double ionization
occurs simultaneously. The solution contains -- besides single ionization waves --
double ionization waves, see Figure \ref{fig:single_and_double_sol}(b). 
These are waves that describe a 2D quantum mechanical system that
is fully broken up into all its sub-particles. In this case, both
relative coordinates $x$ and $y$ can become large, resulting in a
2D scattered wave that extends over to the entire domain.

Note that the principles outlined above can be directly extended to the
three-particle system \eqref{eq:schrod2} with $d = 3$. For energies $E>0$ 
the solution then contains triple ionization waves in addition to the single 
and double ionization waves described above, which simulate the full break-up 
of the 3D partial wave system.

%---------------- SUBSECTION --------------------------------

\subsection{The far field map: classical computation and properties}\label{sec:ffmap}

In physical experiments differential cross sections are measured by detecting 
the fragments that emerge from the break-up problem. These
cross sections are probability distributions describing the
probability of the particle emerging with a certain energy in a certain direction after the
break-up.  Since in practical applications the distance from the target
to the detector is much larger than the size of the system,
the measured cross sections are in fact far field maps of the
scattered wave solution. The numerical calculation of the far field map 
is a two-step process, as outlined in detail below.  

\textbf{Step 1.} First the Schr\"odinger equation \eqref{eq:schrod2} -- 
or equivalently the Helmholtz equation \eqref{eq:helmholtz_ddim} -- is
discretized, meaning the solution is represented on a finite numerical
grid and the differential operators are represented by algebraic
matrix operators. Unless explicitly stated, we generally use 
an equidistant, dimension-uniform discretization, 
where the number of grid points per spatial dimension is denoted 
by $n_x$, $n_y$, $\ldots$, respectively.  The infinite domain $\mathbb{R}^d$ is replaced by a
finite domain $\Omega \subset \mathbb{R}^d$ using absorbing boundary conditions or absorbing layers,
implemented using e.g.\ Perfectly Matched Layers (PML)
\cite{berenger1994perfectly} or Exterior Complex Scaling (ECS)
\cite{mccurdy2004solving}. For a total of $N = n_x \times n_y$ grid points, 
the discretization of the Schr\"odinger equation \eqref{eq:schrod2} results 
in a large and sparse linear system of
the form
\begin{equation}\label{eq:discrete_schrodinger}
  \sum_{j=1}^N \left(-\frac{1}{2}\left(\Delta^h\right)_{ij} + \left(V\right)_{ij} - E \delta_{ij}\right) u^N_j = f_i, \quad i = 1,\ldots,N,
\end{equation}
where $\left(\Delta^h\right)_{ij}$ is the discrete representation of
the Laplacian, which is sparse and typically depends on a
discretization parameter $h$, $\left(V\right)_{ij}$ is the
representation of the potential, which is often diagonal, and $u_j$
and $f_i$ are the grid representations of the solution and the
right-hand side respectively.  The construction of $\left(\Delta^h\right)$ in the
presence of an absorbing boundary layer is discussed in
\cite{baertschy2001electron}.  In the first step of the classical far
field map calculation the linear system
\eqref{eq:discrete_schrodinger} is then solved resulting in the
numerical solution $u^N$.  Due to the incorporation of absorbing
boundary conditions the system may no longer be Hermitian but rather
complex symmetric. The spectral properties of the operator $\left(\Delta^h\right)$
for a finite difference discretization with an ECS absorbing layer are
discussed in Reps et al.~\cite{reps2010indefinite}, and are further
commented on in Section \ref{sec:spectral} of this paper.  The
resulting linear system \eqref{eq:discrete_schrodinger} is indefinite
and is hence very hard to solve using iterative methods such as
multigrid preconditioned Krylov subspace methods, the traditional
method of choice for large and sparse linear systems
\cite{ernst2010difficult}.  

%% In 2006, the so-called complex shifted
%% Laplacian (CSL) preconditioner was presented by Erlangga et
%% al.\ \cite{erlangga2006novel}, which allows for a more efficient and
%% scalable iterative solution to the indefinite problem. However,
%% perfect Krylov scalability in function of the wavenumber or energy is
%% generally unachievable.

\textbf{Step 2.} The second step extracts the far field map 
from the numerical scattered wave solution $u^N$ to equation \eqref{eq:discrete_schrodinger}. 
Indeed, solving \eqref{eq:discrete_schrodinger} 
yields the numerical solution inside a bounded discretization box. 
However, in practice only the probability distribution
far away from the object is measured. An additional post-processing step is hence
required to translate the numerical solution of Step 1 into the far field map.
The extraction of the far-field map is explained on the continuous
equation.  Splitting the potential $V$ in one-body and two-body
operators, see \eqref{eq:potential_split}, allows us to rewrite the Schr\"odinger equation 
\eqref{eq:schrod2} as
\begin{equation*}
  \left(-\frac{1}{2}\Delta + \bar{V}_1(\x) - E\right) u(\x) =  f(\x) - \bar{V}_2(\x) u(\x),  \quad \text{~for~} \x \in \mathbb{R}^d.
\end{equation*}
For separable potentials such as the one-body potentials, an 
explicit analytical expression exists which satisfies 
 $ \left( -\frac{1}{2}\Delta  + \bar{V}_1(\x) -E\right) G(\x,\y) = \delta(\x-\y) $.
The fundamental solution $G(\x,\y)$ is called the Green's function, 
and it allows to write the solution $u$ in any point $\x \in \mathbb{R}^d$ as an 
integral %over the numerical domain 
\begin{equation} \label{eq:inteq0}
   u(\x) =  \int_{\mathbb{R}^d} G(\x,\y)\left( f(\y) - \bar{V}_2(\y) u(\y) \right) d \y, \quad \text{~for~} \x \in \mathbb{R}^d.
\end{equation}
This integral can effectively be calculated by substituting
the numerical solution $u^N$ to the linear system
\eqref{eq:discrete_schrodinger} obtained in Step 1 into the
integrand in \eqref{eq:inteq0}, i.e.
\begin{equation} \label{eq:inteq}
   u(\x) =  \int_{\Omega} G(\x,\y)\left( f(\y) - \bar{V}_2(\y) u^N(\y) \right) d \y, \quad \text{~for~} \x \in \mathbb{R}^d.
\end{equation} 
This approach has been successfully used to solve challenging break up problems, 
see \cite{colton1998inverse,kirsch1996introduction,mccurdy2004solving}.

\textbf{Remark 1.} Note that in the above exposition on the far field map 
it is assumed that both the source term $f(\y)$ and the potential $\bar{V}_2$ 
are compactly supported. This assumption is used when computing the numerical 
solution $u^N$ to the system \eqref{eq:discrete_schrodinger} on a bounded numerical 
domain $\Omega$ that covers the support of $f$ and $\bar{V}_2$, and consequently 
using this solution in the restriction of the integral \eqref{eq:inteq} to the 
domain $\Omega$. 

%---------------- SECTION --------------------------------
\section{The complex contour approach for far field map calculation} \label{sec:compl}
  
The traditional far field map calculation is generally not scalable to
very large, high-dimensional multiparticle problems due to the main
computational bottleneck that lies in solving the linear system
\eqref{eq:discrete_schrodinger}.  This system has similar properties
to the indefinite Helmholtz problem that appears in e.g.\ seismic
inversion \cite{riyanti2006new}. The construction of efficient solvers
for the indefinite Helmholtz system is still an open and active
research topic within the applied mathematics community. However,
since we are in fact not interested in the numerical solution $u^N$ to
\eqref{eq:discrete_schrodinger} itself, but rather in the far field
map of this solution, i.e.\ a Green's integral over $u^N$, the actual
path of integration can be deformed. This gives additional freedom in
the development of the solution method.

%---------------- SUBSECTION --------------------------------
\subsection{Deforming the far field integral}

\begin{figure}
\begin{center}
\includegraphics[width=0.45\textwidth]{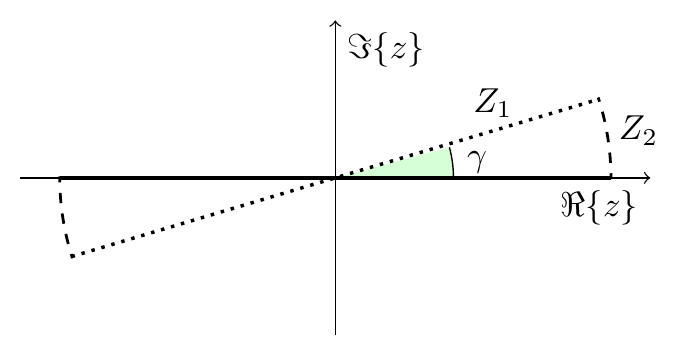} \vspace{-0.3cm}
\caption{Schematic representation of the complex contour for the far field integral calculation \eqref{eq:inteq2} illustrated in 1D. The full line represents the real-valued computational domain $\Omega$, the dotted and dashed lines represent the complex-valued subdomains $Z_1=\{ \x e^{i\gamma} : x \in \Omega \subset \mathbb{R} \}$ and $Z_2= \{b e^{i\theta} : b \in \partial\Omega,~ \theta \in [0,\gamma]\}$ of the complex contour.}
\label{fig:compcont}
\end{center}
\end{figure}

In \cite{cools2014efficient} an efficient method for the computation
of the far field map for indefinite Helmholtz problems was proposed.
It was suggested that if $f$ and $\bar{V}_2$ are analytical functions, 
the integral in equation \eqref{eq:inteq} can alternatively be evaluated 
along a complex contour $Z \subset \mathbb{C}^d$, i.e.
\begin{equation} \label{eq:inteq2}
   u(\x) = \int_{Z} G(\x,\z)\left( f(\z) - \bar{V}_2(\z) u(\z) \right) d\z,
\end{equation}
rather than the physical real-valued domain $\Omega \subset \mathbb{R}^d$. 
The complex-valued domain $Z$ is illustrated schematically for the 1D 
case in Figure \ref{fig:compcont}. The illustrated complex contour grid 
can be readily extended to $d$ dimensions through the use of Kronecker 
products. Note that the integral over the subdomain $Z_2$ vanishes; hence, 
the integral \eqref{eq:inteq2} is in fact only evaluated along the complex 
rotated grid $Z_1=\{ \z = \x e^{i\gamma}: \x \in \Omega \subset \mathbb{R}^d \}$.

The far field map computation from the complex formulation \eqref{eq:inteq2} 
now requires the Schr\"odinger equation \eqref{eq:schrod2} to be solved 
along a complex-valued contour in order to obtain $u(\z)$ for $\z \in Z$.  
It was shown in the literature that the spectral properties of the 
corresponding Schr\"odinger operator discretized along the complex contour 
are favorable for iterative solution \cite{erlangga2006novel,reps2010indefinite}.
Indeed, equation \eqref{eq:schrod2} is reduced to a damped equation along the 
contour, which is much more amenable to iterative solution, particularly using 
multigrid. A key requirement for the validity of the complex contour integral 
reformulation \eqref{eq:inteq2} is the observation that the potentials appearing 
in the context of quantum mechanical break-up problems are continuous functions 
everywhere in the domain except at the exact nuclear positions. This is in 
contrast to many engineering and seismic problems, where the wavenumber (potential) 
often represents objects that exhibit material jumps (discontinuities).

\textbf{Remark 2.} Note that the deformation of the integrating domain \eqref{eq:inteq2}
is only valid for analytical functions $f$ and $\bar{V}_2$, which seemingly contradicts
the earlier assumption that $f$ and $\bar{V}_2$ are compactly supported, see Section 
\ref{sec:ffmap}. The far field map computation in Section \ref{sec:ffmap} can however be 
extended to the more general class of analytical functions $f$ and $\bar{V}_2$ that vanish 
at infinity. Indeed, using smooth bump functions \cite{johnson2007saddle,mead1973convergence}, 
functions with compact support can be shown to be dense within the space of functions that 
vanish at infinity. Consequently, every analytical function can be arbitrarily closely 
approximated by a series of compactly supported functions, which in turn implies that the 
corresponding Helmholtz solutions on a limited computational domain can be arbitrarily close 
to the solution of the Helmholtz equation generated with the analytical vanishing function. 
This means that if $f$ and $\bar{V}_2$ are analytical but sufficiently small everywhere 
outside $\Omega \subset \mathbb{R}^d$, the computational domain for the integral 
\eqref{eq:inteq} may be truncated to the numerical domain $\Omega$ as if $f$ was compactly 
supported. Hence, \eqref{eq:inteq} is well-defined for analytical functions $f$ and $\bar{V}_2$ 
that vanish at infinity, and the deformation to \eqref{eq:inteq2} is valid. 

\textbf{Remark 3.} The supposition that both the source function and the two-body 
potentials are exponentially decaying is a natural assumption in the context of 
photoionization problems. Hence, for photoionization, the complex integral \eqref{eq:inteq2} 
is well-defined and directly yields the ionization cross section. However, when modeling 
electron impact problems, the integral should generally be split into two parts: one real-valued 
integral over the right-hand side $f$ along the real domain $\Omega$, and one complex-valued 
integral over the scattered wave along the complex domain $Z$, i.e.
\begin{equation} \label{eq:inteq3}
   u(\x) = \int_{\Omega} G(\x,\y) \, f(\y) \, d\y - \int_{Z} G(\x,\z) \, \bar{V}_2(\z) u(\z) \, d\z.
\end{equation}
Note that for the left-most integral, the integrand is known explicitly.
The motivation for this splitting originates from the fact that for electron impact models, 
like the 2D Temkin-Poet model (see Section \ref{sec:imple}), the incoming wave $u_{\text{in}}(x,y) = 
\phi_n(x) \sin(k_n y)$ acts as a source term in the right-hand side $f(x,y) = \left( V_1(x)+V_{12}(x,y) \right) u_{\text{in}}(x,y)$. 
Here $k_n = \sqrt{2(E-\lambda_n)}$ and $\phi_n$ is an eigenstate of the one-body operator 
\eqref{eq:1d_eigenstate} with negative eigenvalue $\lambda_n$. The function $\sin(k_n y)$ 
is exponentially increasing along the complex contour, whereas the potentials typically
decrease only linearly. This implies the right-hand side $f$ does not vanish at infinity for 
impact ionization problems. Hence the integral over $f$ cannot be evaluated along the 
complex contour and must be treated separately as a real-valued integral instead.

Note that splitting the integral as suggested above has one important technical consequence. 
When summing up the real- and complex-valued integrals in \eqref{eq:inteq3}, the normalization for the 1D
eigenstates $\phi_n$ has to be chosen consistently.\footnote{For completeness, note that the 
eigenstates $\phi_n$ are a factor of $u_{in}$ in the real-valued part of the integral \eqref{eq:inteq3}; 
however, $\phi_n$ also intrinsically appears in the complex-valued integral over the scattered wave as part of 
the right-hand side of equation \eqref{eq:discrete_schrodinger}.} 
Hence, we propose to use the alternative normalization $ \int_0^{\infty} \phi(z)^2 dz= 1 $ 
for the eigenstates $\phi$ in this work, rather than the more commonly used Hermitian inner 
product with complex conjugation. In this way the complex-valued eigenmodes are normalized 
identically to the normalization on the real-valued domain where the Hamiltonian is Hermitian.

%Since $f(\y)$ and $\bar{V}_2(\y)$ are approximately zero outside the numerical 
%domain, the integral \eqref{eq:inteq} can thus be replaced by a finite integral over the 
%numerical domain $\Omega \subset \mathbb{R}^d$ that is used to compute the scattering solution in Step 1.

\subsection{The complex contour approach: proof of concept}

We briefly recapitulate some of the key results from our previous work \cite{cools2014efficient}, where the 
applicability of the complex contour method on 2D and 3D Schr\"odinger equations
describing a quantum mechanical scattering problem was validated. 

\begin{figure}[t!] 
\begin{center}
\includegraphics[width=0.45\textwidth]{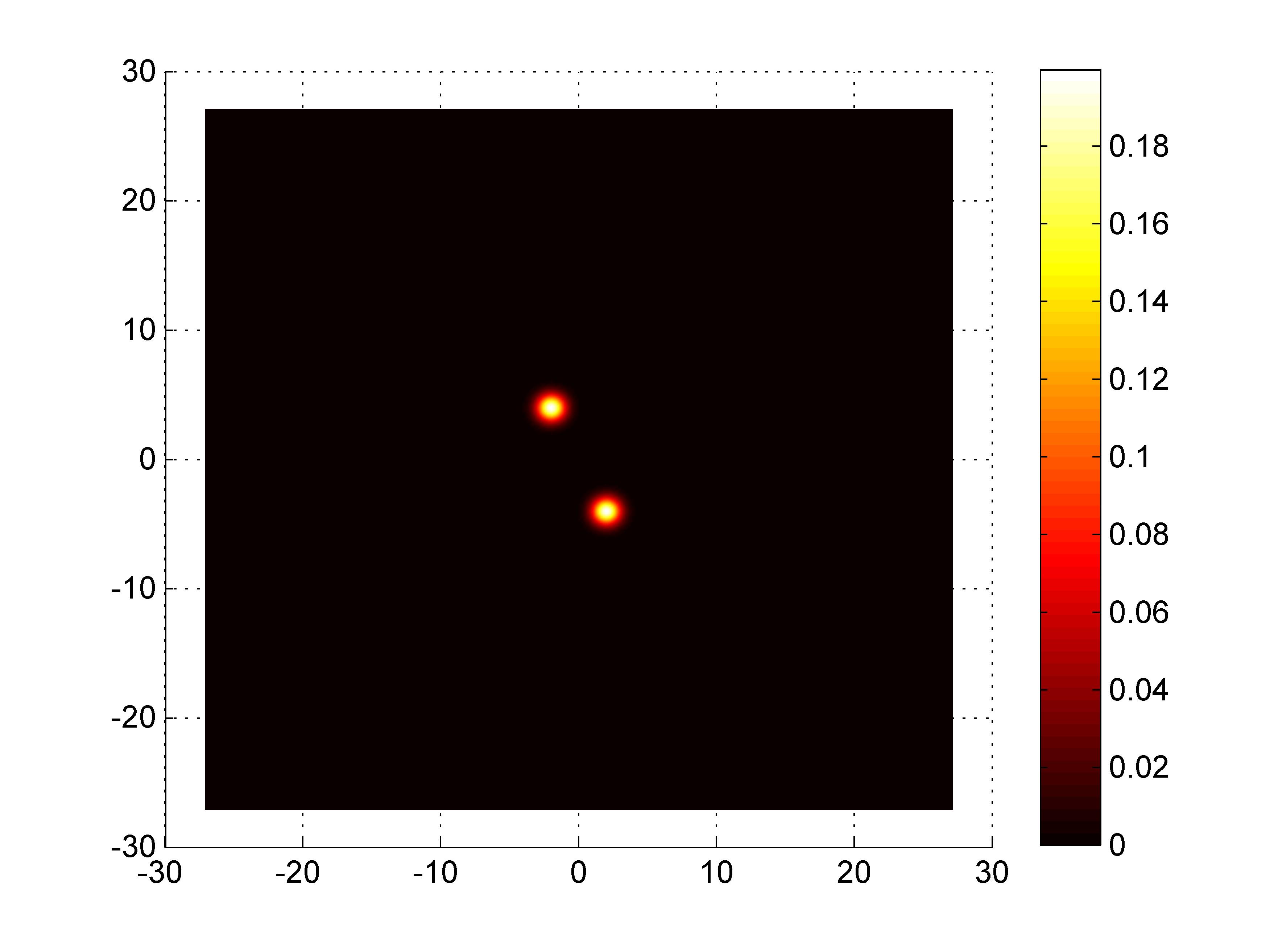}
\includegraphics[width=0.45\textwidth]{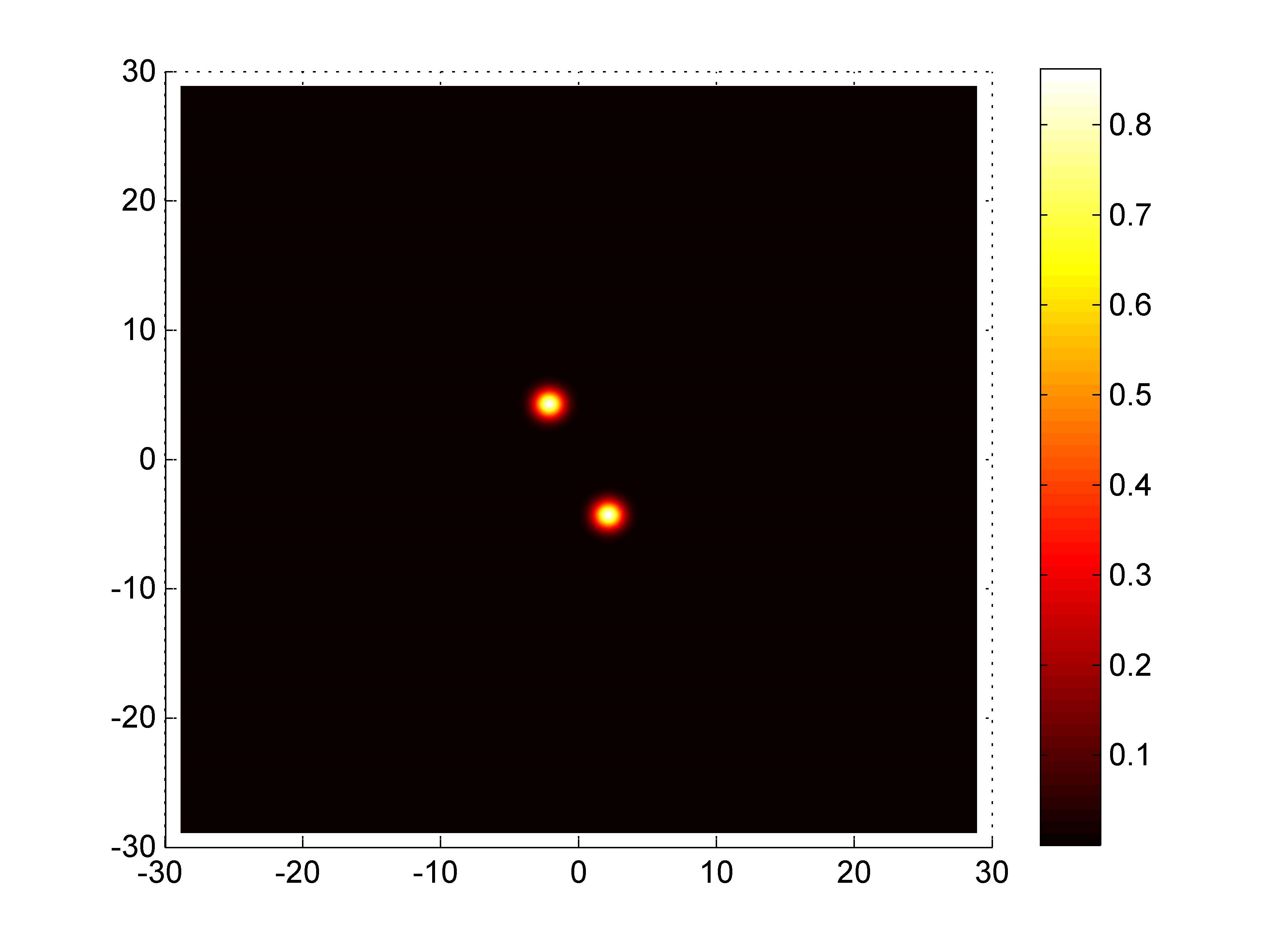}\\ 		\vspace{0.0cm}
\includegraphics[width=0.45\textwidth]{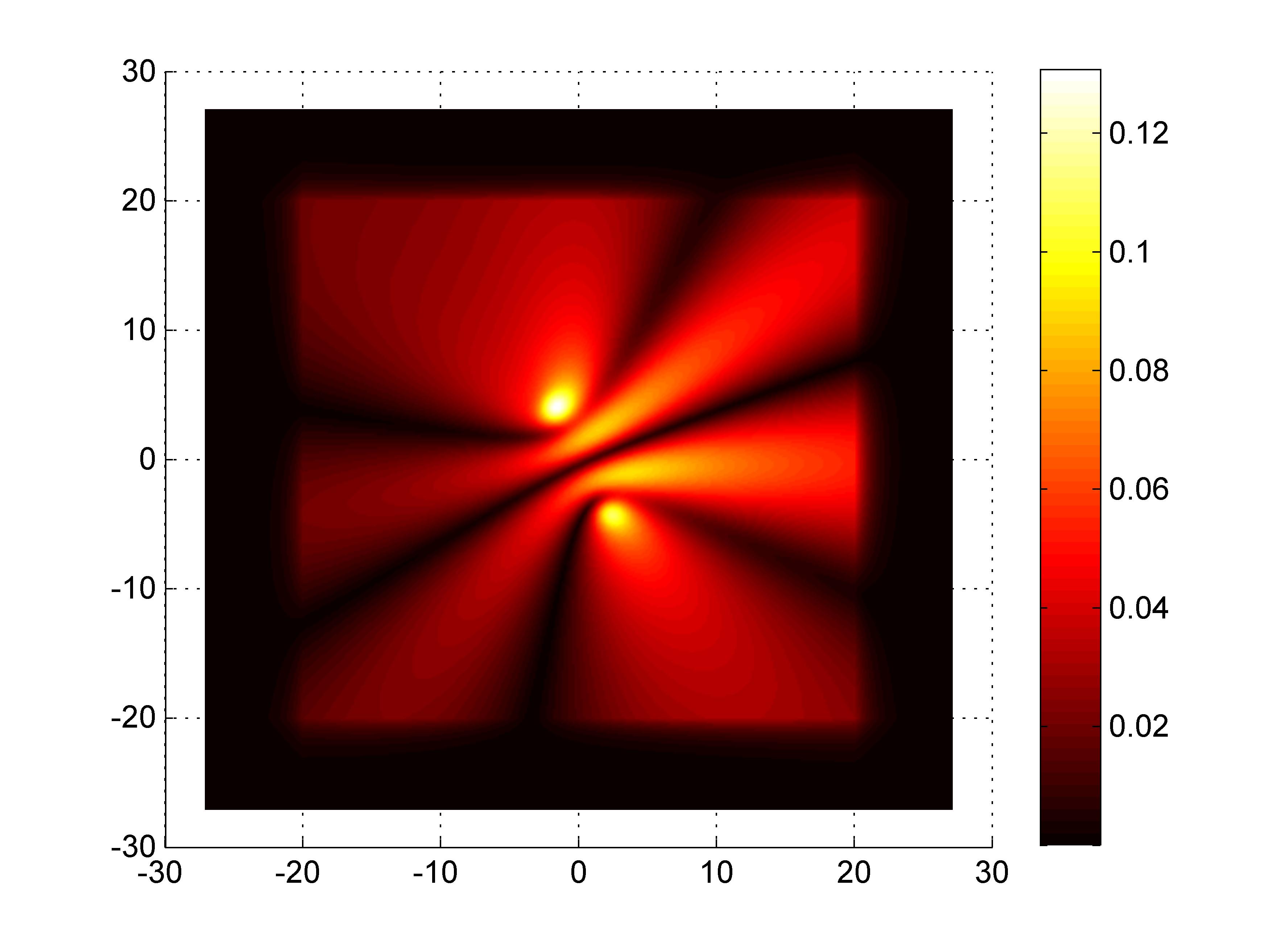}
\includegraphics[width=0.45\textwidth]{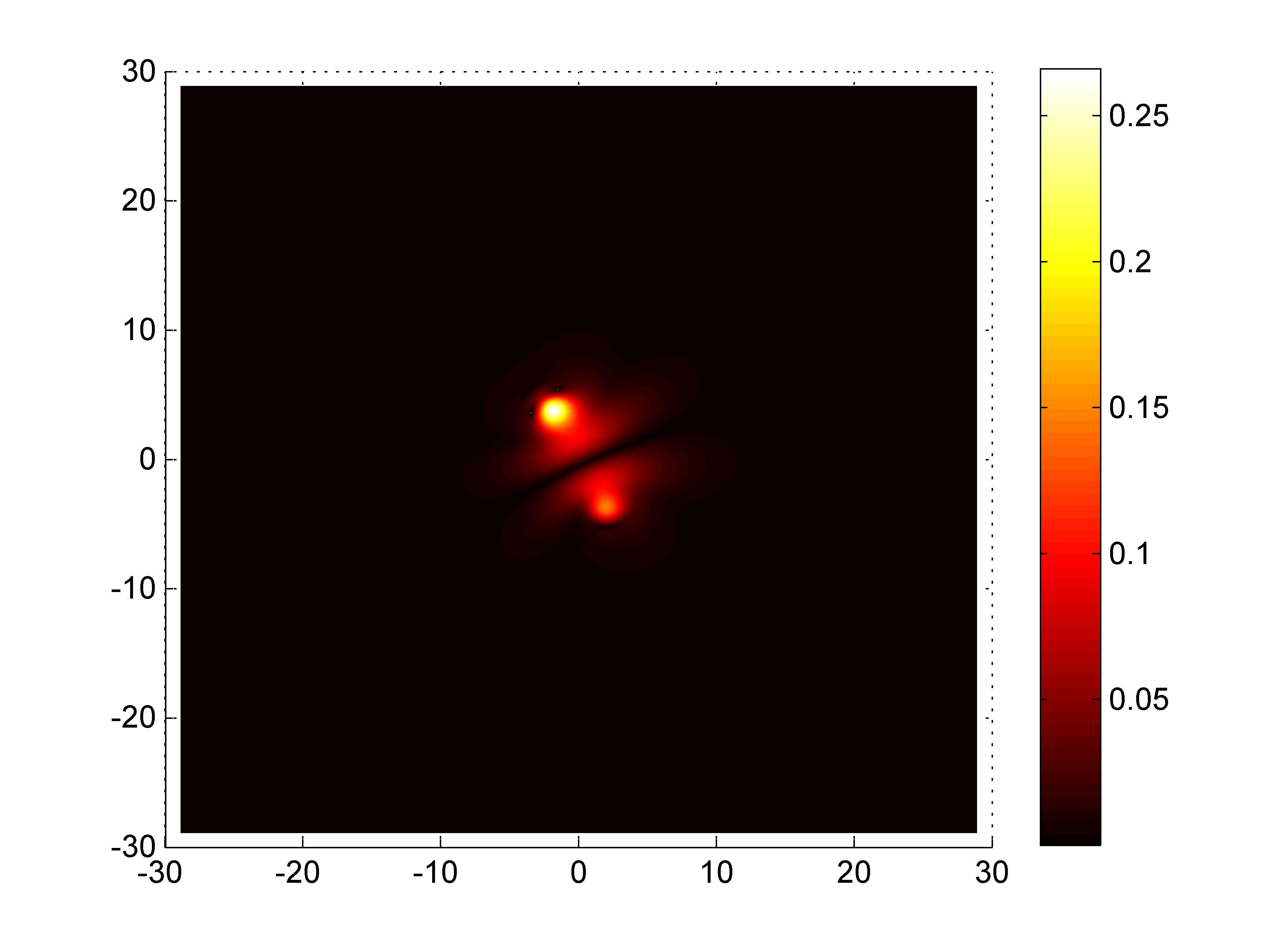}\\ 		\vspace{0.0cm}
\includegraphics[width=0.45\textwidth]{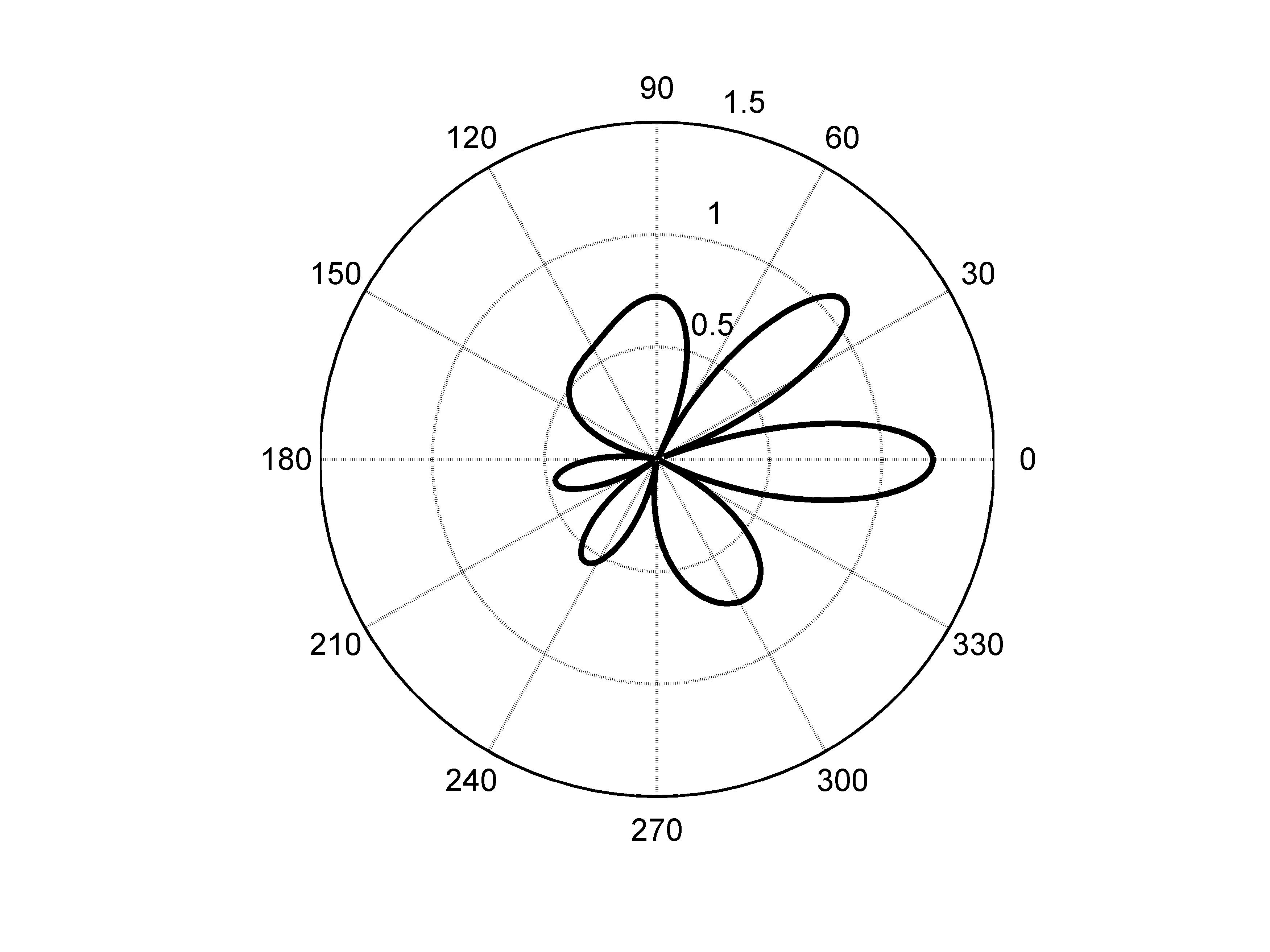}
\includegraphics[width=0.45\textwidth]{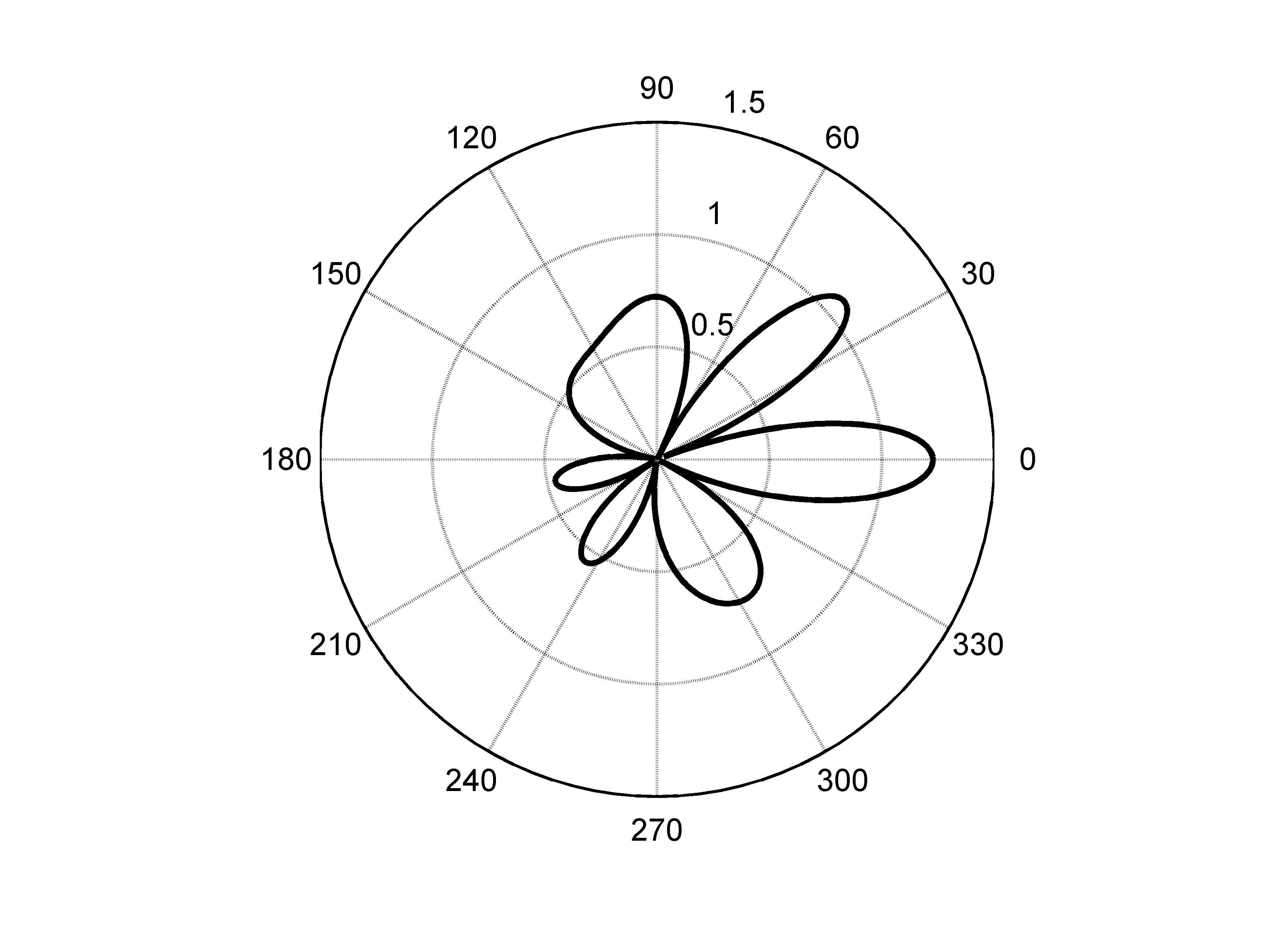} 	  \vspace{-0.4cm}
\caption{Comparison between the classical real-valued far field map calculation (left) and the complex-valued approach (right). Top: 2D object contrast function $|k^2(\x)-k_0^2|$ given by \eqref{eq:chi2D}. Mid: solutions to the Helmholtz problem \eqref{eq:helmholtz_ddim} (modulus) on an $n_x \times n_y = 256 \times 256$ grid. Real-valued problem (incl.~double ECS contour with $\theta_{ECS} = \pi/4$) solved using LU factorization; complex-valued problem solved using a series of multigrid V-cycles with $\omega$-Jacobi smoother on a full complex contour with $\gamma = 14.6^\circ$ (residual reduction tolerance \texttt{1e-6}). Bottom: 2D Far field maps $F(\boldsymbol\alpha)$ for both approaches.}
\label{fig:results2D}
\end{center}
\end{figure}

Consider a 2D $(d=2)$ Helmholtz equation of the form \eqref{eq:helmholtz_ddim} 
with source function $g(\x) = (k^2(\x)-k_0^2) \, u_{\text{in}}(\x)$. 
The equation is discretized using second order finite differences
on an $n^d$-point uniform Cartesian mesh covering a square numerical domain 
$\Omega = [-20,20]^d$. The space-dependent wavenumber is 
\begin{equation} \label{eq:chi2D}
	k^2(x,y) = k_0^2 - 1/5 \left( e^{-((x+2)^2+(y-4)^2)} + e^{-((x-2)^2+(y+4)^2)} \right), \quad (x,y) \in [-20,20]^2,
\end{equation}
i.e.\ the object of interest takes the form of two circular point-like
objects with mass concentrated at the Cartesian coordinates $(2,-4)$
and $(-2,4)$, see Figure \ref{fig:results2D} (top panel). 
The incoming wave scattering at the given object
is defined by
\begin{equation*}
	u_{\text{in}}(\x) = e^{i k_0 \boldsymbol\eta \cdot \x}, \quad \x = (x,y) \in \Omega,
\end{equation*}
where $\boldsymbol\eta$ is the unit vector in the $x$-direction.
The far field amplitude is given by the integral
\begin{equation} \label{eq:FFmap}
	F(\boldsymbol\alpha) = \int_{\Omega} e^{-i k_0 \x \cdot \boldsymbol\alpha} \, [g(\x) + (k^2(\x)-k_0^2) \, u^N(\x)] \, d\x,
\end{equation}
where $\boldsymbol\alpha$ is the outgoing wave unit vector.  Figure
\ref{fig:results2D} illustrates the equivalence between the classical
real-valued far field map integral and the complex contour
formulation.  The 2D Helmholtz model problem with wavenumber given by
(\ref{eq:chi2D}) and $k_0 = 1$ is solved for $u^N$ using, respectively,
a standard LU factorization method on the real domain $\Omega$ with
ECS boundary layers ($\theta_{ECS} = \pi/4$) along the domain boundary
$\partial\Omega$, and a series of multigrid V(1,1)-cycles with
$\omega$-Jacobi smoothing on the full complex domain ($\gamma =
14.6^{\circ} $) with a residual reduction tolerance of
\texttt{1e-6}. The standard multigrid intergrid operators used are bilinear interpolation and full weighting restriction. The
moduli of the wavenumber contrast function $k^2(\x)-k_0^2$ (top) and
the resulting solution $u^N$ (mid) are shown on Figure
\ref{fig:results2D} for both methods. Note how the solution $u^N$ on
the full complex contour (right) is heavily damped compared to the
solution on the real domain (left). Consequently, using the numerical
solution $u^N$, the 2D far field map integral \eqref{eq:FFmap} can be
calculated using any numerical integration scheme over the real or
complex domain respectively. The resulting far field map
$F(\boldsymbol\alpha)$ is shown as a function of the outgoing wave
direction $\boldsymbol\alpha$ on Figure \ref{fig:results2D} (bottom).
One observes that the mapping is indeed identical when calculated over
the real-valued (left) and complex-valued (right) domain. However, the
computational cost of the real-domain method for calculation of the
far field map is significantly reduced by the ability to apply a
multigrid method to the complex scaled problem.

%---------------- SUBSECTION --------------------------------
\subsection{The ionization cross sections}

In practice, the far field maps of \eqref{eq:2d_eigenstate} are Green's integrals 
over the solution, see \cite{mccurdy2004solving}. Indeed, for a 2D problem setting 
where $\x = (x,y)$, the single ionization amplitude $s_n(E)$, which represents the 
probability of particle single ionization, is given by
\begin{equation} \label{eq:single-ionization}
  s_n(E) = \int_\Omega  \phi_{k_n}(x) \phi_n(y) \left[f(x,y) -  V_{12}(x,y) u(x,y)\right]  \, dx \, dy,
\end{equation}
where $k_n = \sqrt{2(E-\lambda_n)}$. Here the function $\phi_n$ is a one-body
eigenstate, i.e.\ a solution of equation \eqref{eq:1d_eigenstate} with corresponding eigenvalue
$\lambda_n$, and the function $\phi_{k_n}$ is a regular, normalized solution of the homogeneous Helmholtz equation
\begin{equation}\label{eq:testfunction}
\left(-\frac{1}{2}\frac{d^2}{dx^2} + V_1(x) -\frac{1}{2}k^2\right)\phi_{k} = 0,
\end{equation}
where $k=k_n$ and $\phi_{k_n}$ is normalized by $1/\sqrt{k_n}$. 
Similarly, the double ionization cross section $\zeta(k_1,k_2)$, which measures the 
probability of double ionized particles, is defined by the integral
\begin{equation} \label{eq:double-ionization}
 \zeta(k_1,k_2) =  \int_\Omega  \phi_{k_1}(x)  \phi_{k_2}(y)\left(f(x,y) -  V_{12}(x,y) u(x,y)\right) \, dx \, dy,  \quad x,y \ge 0,
\end{equation}
where both $\phi_{k_1}(x)$ and $\phi_{k_2}(y)$ are solutions to
\eqref{eq:testfunction}, with $k_1 = \sqrt{2E} \sin(\alpha)$ and $k_2
= \sqrt{2E}\cos(\alpha)$, respectively, and $\alpha \in [0,\pi/2]$
such that $k_1^2 + k_2^2 = 2E$.  The total double ionization cross
section is defined as the integral
\begin{equation} \label{eq:total-double-ionization}
  \sigma_{tot}(E) = \int_0^E \sigma(\sqrt{2\epsilon},\sqrt{2(E-\epsilon)}) \, d\epsilon,
\end{equation}
where 
\begin{equation*}
  \sigma(k_1,k_2) = \frac{8\pi^2}{k_0^2} \frac{1}{k_1k_2} |\zeta(k_1,k_2)|^2.
\end{equation*}
Note that the cross section integral expressions \eqref{eq:single-ionization} and 
\eqref{eq:double-ionization} are effectively of the form \eqref{eq:inteq}, i.e.~these are
far field maps of the Schr\"odinger solution $u(x,y)$, and the complex contour approach is hence applicable.

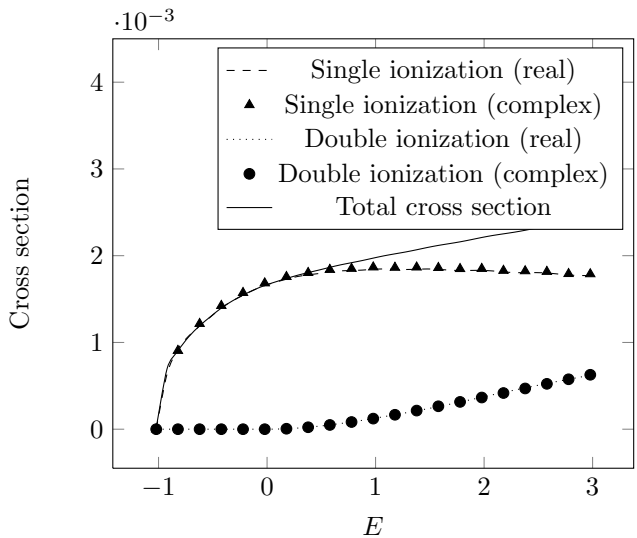
\begin{figure}
\begin{center}
 \input{total} \vspace{-0.3cm}
\end{center}
\caption{Comparison of the single and double ionization total cross
  sections calculated using the numerical scattered wave solution $u^N$ to \eqref{eq:partialwave2d}. Computation performed on (a) a traditional 
  real-valued ECS grid with $\theta_{ECS} = 25.7^\circ$ and (b) a full complex contour with $\gamma = 8.5^\circ$. 
  The energy range starts at the single ionization threshold $E=-1$, corresponding to a strictly positive cross section.
  Double ionization occurs for energy levels $E>0$. 
  \label{fig:single_and_double} }
\end{figure}

Figure \ref{fig:single_and_double} shows the rate of single and double
ionization as a function of the total energy $E$ for a 2D 
Schr\"odinger model of the form \eqref{eq:partialwave2d} 
with source term $f(x,y) = \exp(-3(x+y)^2)$.
The potentials are exponential functions defined as $V_1(x) = -4.5\exp(-x^2)$, 
$V_2(y) = -4.5\exp(-y^2)$ and $V_{12}(x,y) = 2\exp(-(x+y)^2)$.
The dashed and dotted lines on Figure \ref{fig:single_and_double} 
represent the total single and double ionization amplitudes
calculated using the traditional real-valued method with ECS absorbing
boundary conditions \cite{mccurdy2004solving}. 
The corresponding 2D scattering
problem (\ref{eq:partialwave2d}) is solved on a numerical domain
$\Omega = [0,15]^2$ covered by a finite difference grid consisting of
300 grid points per spatial dimension. An ECS
absorbing boundary layer of 150 additional grid points per dimension 
starting at $x = 15$ ($y = 15$ resp.) simulates the outgoing wave boundary 
conditions for $x \to \infty$ and $y \to \infty$ ($\theta_{ECS} = \pi/7 \approx 25.7^\circ$).  
Equivalent results obtained using the complex contour approach are 
indicated by the $\blacktriangle$ and $\bullet$ symbols on Figure 
\ref{fig:single_and_double}. The Schr\"odinger equation \eqref{eq:partialwave2d} is first
solved on a complex contour, yielding a damped solution, followed by the
calculation of the integrals \eqref{eq:single-ionization} and
\eqref{eq:double-ionization} along the complex domain. 
A complex scaled grid with an overall 
complex rotation angle $\gamma = \pi/21 \approx 8.5^\circ$ is used.  

%---------------- SUBSECTION --------------------------------
\subsection{Spectral properties of the 2D Schr\"odinger operator}\label{sec:spectral}

\begin{figure}
\begin{center}
 \includegraphics[width=0.58\textwidth]{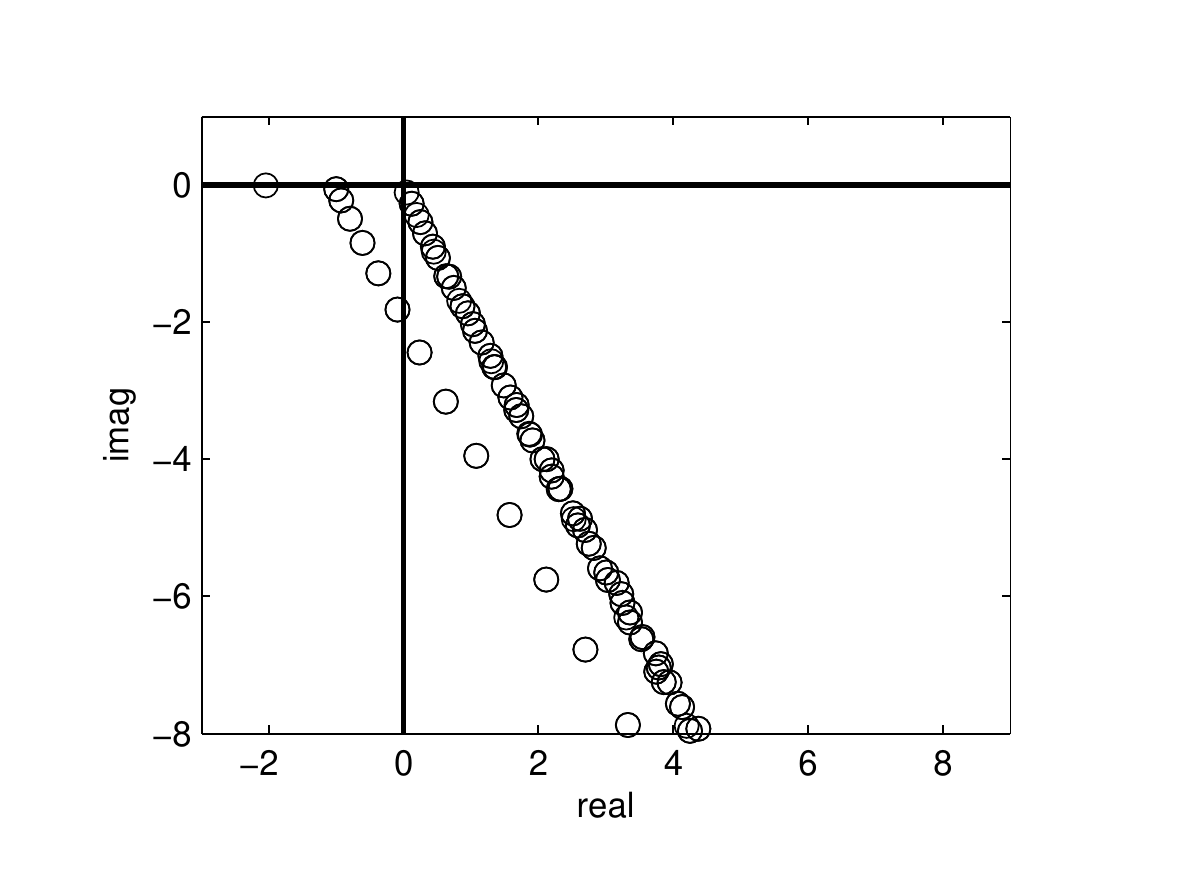} \vspace{-0.5cm}
\end{center}
\caption{Spectrum (close-up) of the discretized 2D Hamiltonian \eqref{eq:hamiltonian2d} 
on a complex contour with grid distance $h e^{-i\gamma}$, where $h = L/n_x = L/n_y$ is 
the corresponding real-valued grid distance with domain length $L$. The complex grid 
rotation angle is $\gamma = \pi/6$, causing the spectrum to be rotated down into the 
complex plane over an angle $2\gamma = \pi/3$.}
\label{fig:eigenvalues}
\end{figure}

To obtain additional insight in the multigrid convergence, 
we briefly discuss the spectral properties of the discretized 2D
Schr\"o\-ding\-er operator. 
The discretized 2D Hamiltonian $H^{2d}$ corresponding to equation
\eqref{eq:partialwave2d} can be written as a sum of two Kronecker
products and a two-body potential, i.e.
\begin{equation} \label{eq:hamiltonian2d}
  H^{\text{2d}} = H_1\otimes I + I \otimes H_2 +  V_{12}(x,y),
\end{equation}
where
\begin{equation*}
H_1 = -1/2 \Delta + V_1 \qquad \textrm{and} \qquad H_2 = -1/2 \Delta + V_2
\end{equation*}
are the discretized one-dimensional Hamiltonians. 
When the two-body potential $V_{12}(x,y)$ is weak relative to the one-body 
potentials $V_1(x)$ and $V_2(y)$, the eigenvalues of the 2D Hamiltonian $H^{2d}$ can be approximated by
\begin{equation*} 
  \lambda^{\text{2d}} \approx \lambda_{1} + \lambda_{2},
\end{equation*}
where $H_1\phi(x) = \lambda_1\phi(x)$ and $H_2 \phi(y) = \lambda_2\phi(y)$.
The spectrum of the one-dimensional Hamiltonians $H_1$ and $H_2$ closely resembles the 
spectrum of the 1D Laplacian $(-1/2) \Delta$; however, the presence of the
potential modifies the smallest eigenvalues, typically yielding a single
(or a small number of) negative eigenvalue(s) due to the attractive potential.

The spectrum of the complex rotated 2D Schr\"odinger operator with
$E=0$ is shown on Figure \ref{fig:eigenvalues}. The eigenvalues of the
2D Hamiltonian are rotated down in the complex plane when the system
is discretized along a complex-valued contour.  An isolated negative
real eigenvalue at $-2.043$ appears, and two series of eigenvalues
emerge from the real axis at $-1.012$ and $0$ respectively. The
leftmost branch of eigenvalues originates from the sum of the negative
eigenvalue of the first 1D Hamiltonian $H_1$ combined with all the
positive eigenvalues of the second 1D Hamiltonian $H_2$, and vice
versa. The second series of eigenvalues starting at the origin is
approximately formed by the sums of the positive eigenvalues of both
one-dimensional Hamiltonians.  The isolated eigenvalue is a bound
state of $H^{2d}$. The leftmost branch correspond to so-called
single-ionization states, i.e.\ highly oscillatory localized 1D
waveforms propagating close to the $x=0$ and $y=0$ axes.

Depending on the sign of the energy $E$, the spectral properties of
the system to be solved change. Its spectrum is then composed of the 
eigenvalues of $H^{2d}$, shown in Fig.~\ref{fig:eigenvalues}, shifted 
to the left or right. For some negative energies $E<0$ (single ionization regime),
this might result in an eigenvalue located close to zero, which causes
most iterative solvers (including multigrid) to suffer from a severe
convergence slow down. This observation forms the major motivation for
the development of an additional correction scheme to support the
multigrid solver. The Coupled Channel Correction Scheme introduced in
the next section is designed specifically to account for the bound
state eigenmodes.

%%%%%%%%%  SECTION  %%%%%%%%%%%%%%%%%%%%%%%%%%%%%%%%%%%%%%%%%%%%%%%%

\section{MG-CCCS: a multigrid-based solver which includes a Coupled Channel Correction Scheme} \label{sec:multi}

It was shown in \cite{cools2014efficient} that multigrid generally
acts as an efficient solution scheme for the discretized Schr\"odinger
system \eqref{eq:discrete_schrodinger} when the latter is represented
along a complex contour.  However, it was also observed that multigrid
convergence tends to stagnate for energies $E<0$. Notably, a
deterioration in convergence occurs when the problem is not fully
broken up in its subparticles, in which case the convergence rate is
strongly affected by the presence of eigenvalues near the real axis on
both sides of the origin, cf.\,Section \ref{sec:spectral}.  

In this section, we present a correction scheme that can be used in addition
to the standard multigrid scheme, and which is designed specifically
at resolving the low-dimensional single ionization waves in the
solution.  Using the Coupled Channel approximation
\cite{heller1973comment,mccarthy1983momentum} we obtain a system of
$(d-1)$-dimensional Coupled Channel equations, which can be solved
directly due to the dimension reduction, and can consequently be used
to correct the solution.  The separate handling and elimination of the
bound states from the error results in an overall more robust
iterative solution scheme.  We derive the Coupled Channel scheme in a
2D problem setting for notational and implementation convenience.

%\subsection{On the normalization of the eigenmodes}
%
%We briefly comment on the normalization of the eigenstates used throughout this section. 
%The traditional normalization for general complex-valued eigenfunctions $\phi$ is given by
%\begin{equation} \label{eq:stand_norm}
%\int_0^{\infty} \phi(z)^* \phi(z) = 1,
%\end{equation}
%where $^*$ denotes the complex conjugation. 
%Indeed, after discretization standard numerical eigenvalue routines typically normalize
%the eigenstates $\phi \in \mathbb{C}^n$ such that $\sum_{i=1}^N \phi_i^*\phi_i =1$. 
%Note however that this leads to a different normalization of the 
%target states before and after complex rotation, which is highly inconvenient. 
%In this work, we therefore use the alternative normalization 
%\begin{equation} \label{eq:our_norm}
%  \int_0^{\infty} \phi(z)^2 dz= 1,
%\end{equation}
%rather than the standard normalization using complex conjugation \eqref{eq:stand_norm}.  
%In this way the complex-valued eigenmodes are normalized identically to the normalization 
%on the real part of the domain, where the Hamiltonian is Hermitian and each eigenstate can 
%hence be described by a real-valued function.

%---------------- SUBSECTION --------------------------------
\subsection{The Coupled Channel approximation}

The aim of the correction scheme is to approximately solve the driven Schr\"odinger equation 
\begin{equation}\label{eq:driven}
  (H-E) \, u(x,y) = ( H_1 + H_2 + V_{12}(x,y) - E) \, u(x,y) = f(x,y)
\end{equation}
for an energy $E$ where single ionization waves dominate the
solution. This implies that, in large parts of the domain, the solution can be
written as the product of an outgoing wave in one coordinate with an eigenstate in the
other coordinate. Let us assume that the 1D Hamiltonians $H_1$ and $H_2$ are
such that there are eigenstates
\begin{equation*}
  H_1 \phi_n(x) = \lambda_n \phi_n(x), 
\end{equation*}
\begin{equation*}
  H_2 \varphi_n(y) = \mu_n \varphi_n(y), 
\end{equation*}
with $\lambda_n<0$ and $\mu_n<0$. 
Using the so-called Coupled Channel approximation, 
we approximate a given function $u(x,y)$ as
\begin{equation}\label{eq:coupled_channel_expansion}
  u(x,y) \approx \sum_{m=1}^{M} A_m(y) \phi_m(x)  + \sum_{l=1}^L B_l(x) \varphi_l(y),
\end{equation}
where $M$ and $L$ are assumed to be much smaller that the number of discretization
points in each respective spatial dimension.  The coefficient functions $A_m$ and $B_l$ can represent, for example,
the asymptotic form of a single ionization outgoing wave, 
for which it is known that $\lim_{y \rightarrow \infty} u(x,y) = \sum_{m=1}^M S_m \exp(i k_m y) \phi_m(x)$,
where $S_m$ are the single ionization amplitudes and $k_m = \sqrt{2(E-\lambda_m)}$.

\textbf{Remark 4.}
  Note that the approximate expansion \eqref{eq:coupled_channel_expansion} is generally not
  uniquely defined. Indeed, we can add arbitrary linear combinations of the 
  eigenfunctions $\varphi_l(y)$ to $A_m(y)$ without modifying the approximation. 
  For example, adding $\sum_{l=1}^L \alpha_{lm} \varphi_l(y)$ to $A_m(y)$ is compensated 
  by subtracting a similar linear combination of the eigenmodes $\phi_m(x)$ from $B_l(x)$, 
  i.e.
\begin{align*}
\sum_{m=1}^{M} \left(A_m(y) + \sum_{l=1}^L \alpha_{lm} \varphi_l(y)\right) \phi_m(x)  + \sum_{l=1}^L \left(B_l(x)-\sum_{m=1}^M \alpha_{lm} \phi_m(x)\right)  \varphi_l(y) \\
=\sum_{m=1}^{M} A_m(y) \phi_m(x) + \sum_{l=1}^{L} B_l(x) \varphi_l(y).
\end{align*}
To define a unique representation of the form \eqref{eq:coupled_channel_expansion}, 
we choose, without loss of generality, $A_m(y)$ and $B_l(x)$ such that (for $1 \leq i \leq M$, $1 \leq j \leq L$)
\begin{equation}\label{eq:convention_a}
a_{ji} :=  \int_0^\infty  A_i(y) \varphi_j(y) dy =  0, \qquad \forall j > i,
\end{equation}
\begin{equation}\label{eq:convention_b}
b_{ij} :=  \int_0^{\infty} B_j(x) \phi_i(x) dx=  0, \qquad \forall  i \geq j .
\end{equation}
Note, however, that any other convention, for example that all $a_{ji} =0$ for all $i,j$, would also be possible.\vspace{0.3cm} \\
Given a known function $u(x,y)$ we can now calculate the corresponding coefficient functions $A_m(y)$ and $B_l(x)$ 
of the Coupled Channel approximation as follows. For $1 \leq i \leq M$ and $1 \leq j \leq L$, it holds that
\begin{equation*}
  A_i(y) = \int_0^\infty \hspace{-0.1cm} u(x,y) \phi_i(x) dx - \sum_{l=1}^L \left(\int_0^\infty B_l(x) \phi_i(x) dx\right) \varphi_l(y) = \int_0^\infty \hspace{-0.1cm} u(x,y) \phi_i(x) dx - \sum_{l=1}^L b_{il}\varphi_l(y),  
\end{equation*}
\begin{equation*}
  B_j(x) = \int_0^\infty \hspace{-0.1cm} u(x,y) \varphi_j(y) dy - \sum_{m=1}^M \left(\int_0^\infty A_m(y) \varphi_j(y) dy\right) \phi_m(x) = \int_0^\infty \hspace{-0.1cm} u(x,y) \varphi_j(y) dy - \sum_{m=1}^M a_{jm}\phi_m(x).   
\end{equation*}
Hence, the coefficients $a_{ji}$ and $b_{ij}$ need to be computed in order to calculate $A_i(y)$ and $B_j(x)$.
These are found directly from  $u(x,y)$ by integration, since
\begin{equation} \label{eq:coefficients}
  \int_0^\infty \int_0^\infty \phi_i(x) \varphi_j(y) \, u(x,y) \, dx dy = a_{ji} + b_{ij}. 
\end{equation}
Since $b_{ij}$ is zero while $a_{ji}$ is non-zero and vice versa due the uniqueness 
conventions \eqref{eq:convention_a}-\eqref{eq:convention_b}, the coefficients $a_{ji}$ 
and $b_{ij}$ are indeed uniquely defined by \eqref{eq:coefficients}, and hence the 
expansion \eqref{eq:coupled_channel_expansion} can be computed.

%---------------- SUBSECTION --------------------------------
\subsection{The Coupled Channel projection space}

Given the 2D driven Schr\"odinger equation, we now represent 
both the unknown solution and the known right-hand side in a factorized way using the 
Coupled Channel approximation \eqref{eq:coupled_channel_expansion}. 
Equation \eqref{eq:driven} becomes 
\begin{equation*}
(H_1+ H_2 + V_{12}(x,y) -E) \left(\sum_{m=1}^{M} A_m(y) \phi_m(x) + \sum_{l=1}^L B_l(x) \varphi_l(y) \right)= \sum_{m=1}^M f_m^A(y) \phi_m(x) + \sum_{l=1}^L f^B_l(x)\varphi_l(y),
\end{equation*}
where $f_m^A$ and $f_l^B$ are uniquely defined, cf.\,\eqref{eq:convention_a}-\eqref{eq:convention_b}, and can be calculated explicitly. A decomposition of this equation gives rise to a pair of coupled systems of equations. The first system is found by starting from
\begin{align*}
  \left(H_1 + H_2 + V_{12}(x,y) -E\right) \left(\sum_{m=1}^M A_m(y) \phi_m(x) \right) = \sum_{m=1}^M f^A_m(y) \phi_m(x).
\end{align*}
Using the fact that $H_1\phi_n(x)  = \lambda_n \phi_n(x)$, multiplying the above 
equation with $\phi_i(x)$ (for $1\leq i \leq M$) and integrating over $x$ yields
\begin{equation*}
  (H_2 + \lambda_i-E) \, A_i(y) + \sum_{m=1}^M \left(\int_0^\infty V_{12}(x,y) \phi_i(x) \phi_m(x) dx\right) A_m(y) = f^A_i(y).
\end{equation*}
Note that we use a specific normalization of the eigenmodes in order to obtain the above equation.\footnote{Throughout 
this work we use an alternative normalization of the complex-valued eigenmodes $\phi$ given by
$ \int_0^{\infty} \phi(z)^2 dz= 1 $,
rather than the more commonly used Hermitian inner product with complex conjugation. See also Remark 3, Section \ref{sec:compl}.}
Introducing the notation $V^A_{im}(y) := \int_0^{\infty} V_{12}(x,y)\phi_i(x)\phi_m(x) dx$ ($1\leq i,m \leq M$), 
this ultimately results in the Coupled Channel equation for $A_i(y)$
\begin{equation} \label{eq:channel1}
\left(H_2 + \lambda_i + V^A_{ii}(y) - E\right) A_i(y) + \sum_{\substack{m=1\\m\neq i}}^M V^A_{im}(y) A_m(y) = f^A_i(y) \quad \text{for}\quad i=1,\ldots,M.
\end{equation}
In a similar way, one can derive a coupled system for $B_j(x)$
\begin{equation} \label{eq:channel2}
\left(H_1 + \mu_j + V^B_{jj}(x) - E\right) B_j(x) + \sum_{\substack{l=1\\l\neq j}}^L V^B_{jl}(y) B_l(x) = f^B_j(y) \quad \text{for}\quad j=1,\ldots,L,
\end{equation}
where $V^B_{jl}(y) := \int_0^{\infty} V_{12}(x,y)\varphi_j(y)\varphi_l(y) dy$  ($1\leq j,l \leq L$). 
These equations are known as the Coupled Channel equations and are frequently
used to describe electronic states in atoms and molecules, see \cite{friedrich2006theoretical}.

For the 2D Schr\"odinger system \eqref{eq:driven} the corresponding Coupled Channel 
equations \eqref{eq:channel1}-\eqref{eq:channel2} form a set of $M+L$ one-dimensional
 equations that need to be solved in order to obtain the Coupled Channel representation 
 \eqref{eq:coupled_channel_expansion} for the solution $u(x,y)$. Hence, the Coupled Channel approximation can be considered 
 as a dimension reducing projection of the 2D solution $u(x,y)$ onto a collection of 1D subproblems.
 
%---------------- SUBSECTION --------------------------------
\subsection{The Coupled Channel Correction Scheme}

We now use the Coupled Channel equations \eqref{eq:channel1}-\eqref{eq:channel2} to define 
a correction scheme for the solution of the driven Schr\"odinger equation.
Let $u^{(k)}(x,y)$ be a $k$-th guess for the solution of equation \eqref{eq:driven}, 
which was computed by some iterative solution scheme (e.g.\ a series of multigrid V-cycles)
from a previous guess $u^{(k-1)}(x,y)$. 
An improved guess $u^{*(k)}$ can be obtained with the help of the following correction scheme. 
We first formulate the error equation
\begin{equation*}
  (H-E) \, e^{(k)}(x,y) = r^{(k)}(x,y),
\end{equation*}
where $e^{(k)}(x,y) = u(x,y)-u^{(k)}(x,y)$ is the current guess error and $r^{(k)} = f(x,y) - (H-E)\,u^{(k)}(x,y)$
is the corresponding residual. Instead of solving the error equation exactly, which is as hard as solving the 
original problem, we solve it approximately using the Coupled Channel approximation \eqref{eq:coupled_channel_expansion}.  
We expand the residual and error functions using the Coupled Channel approximation, such that
\begin{equation*}
r^{*(k)}(x,y) = \sum_{m=1}^M r_m^A(y) \phi_m(x) + \sum_{l=1}^L r^B_l(x)\varphi_l(y),
\end{equation*}
\begin{equation*}
e^{*(k)}(x,y) = \sum_{m=1}^M e_m^A(y) \phi_m(x) + \sum_{l=1}^L e^B_l(x)\varphi_l(y).
\end{equation*}
Here, the coefficient functions $r_m^A(y)$ and $r^B_l(x)$ for the residual are known.
Subsequently, the Coupled Channel equations are solved for the error coefficient functions $e^A_i(y)$ and $e^B_j(x)$:
\begin{equation*}
\left(H_2 + \lambda_i + V^A_{ii}(Y) - E\right) e^A_i(y) + \sum_{\substack{m=1\\m\neq i}}^M V_{im}(y) e^A_m(y) = r^A_i(y) \quad \text{for}\quad i=1,\ldots,M,
\end{equation*}
\vspace{-0.3cm}
\begin{equation*}
\left(H_1 + \mu_j + V^B_{jj}(x) - E\right) e^B_j(x) + \sum_{\substack{l=1\\l\neq j}}^L V^B_{jl}(y) e^B_l(x) = r^B_j(y) \quad \text{for}\quad j=1,\ldots,L.
\end{equation*}
The solution of these one-dimensional systems is significantly easier 
than solving the original 2D driven Schr\"odinger system \eqref{eq:driven}.
Indeed, the reduction in spatial dimension allows us to effectively solve the Coupled Channel 
systems using a direct solution method, which yields a fast and accurate solution.
We then correct the current guess for the solution by adding the Coupled Channel 
approximation of the error $e^{*(k)}(x,y)$ to the current solution $u^{(k)}(x,y)$ as follows
\begin{align}
  u^{*(k)}(x,y) &= u^{(k)}(x,y) + e^{*(k)}(x,y) \notag \\
  							&= u^{(k)}(x,y) + \sum_{m=1}^{M} e^A_m(y) \phi_m(x) + \sum_{l=1}^L e^B_l(x) \varphi_l(y). \label{eq:cccs}
\end{align}

\textbf{Remark 5.} Note that the extension of the above Coupled Channel Correction
Scheme \eqref{eq:cccs} to higher spatial dimensions, notably 3D, can be
obtained in an analogous manner. Such a 3D correction scheme would be
based upon a double Coupled Channel approximation, introducing
corrections for both the 1D and 2D eigenstates to account
for the presence of single and double ionization waves travelling
along the axes and faces of the coordinate system, respectively. This
leads to both 1D and 2D pairs of coupled systems of equations, which
are in turn significantly easier to solve than the full 3D break-up
problem.  Although we believe the generalization to higher dimensions
is straightforward, we do not expound on the 3D Coupled Channel
Correction Scheme in this work for notational and implementation ease.
Instead, we aim to tackle this problem in future work.

%%%%%%%%%  SECTION  %%%%%%%%%%%%%%%%%%%%%%%%%%%%%%%%%%%%%%%%%%%%%%%%

\section{Implementation and numerical accuracy} \label{sec:imple}

Additional care is required for the calculation of the eigenstates
involved in the scattering process when the coordinates are rotated 
into the complex plane. For photoionization problems, the 
right-hand side is typically the dipole operator applied to the eigenstate 
with the lowest energy \cite{vanroose2006double}. However, for impact 
ionization problems the right-hand side contains the incoming wave that
consists of a target state, i.e.\ the eigenstate with the lowest energy, 
and a purely incoming wave representing the incoming particle 
\cite{baertschy2001electron,mccurdy2004solving}. Both of these functions require a careful 
and accurate numerical description, especially on the complex contour, 
since after complex rotation the incoming wave becomes an exponentially 
increasing function.

%---------------- SUBSECTION --------------------------------
\subsection{The complex contour rotation angle and numerical accuracy} 

In this section we comment on some key numerical aspects of the proposed
complex contour approach. From a theoretical viewpoint the real-valued 
integral \eqref{eq:inteq} and its counterpart evaluated 
along the complex contour \eqref{eq:inteq2} are identical, independently 
of the size of the rotation angle $\gamma$. Indeed, in principle the
rotation angle can be chosen arbitrarily large, implying very strong 
damping, which is beneficial for the multigrid solver.
However, in reality a restriction on the size of the rotation angle
is imposed by the numerical implementation.
We illustrate this claim on a 2D Temkin-Poet model problem, 
where it is shown that when the angle of rotation is too large and the 
potential is long range, significant numerical inaccuracies occur.

\begin{figure}
\begin{center}
  \includegraphics[width=0.5\textwidth]{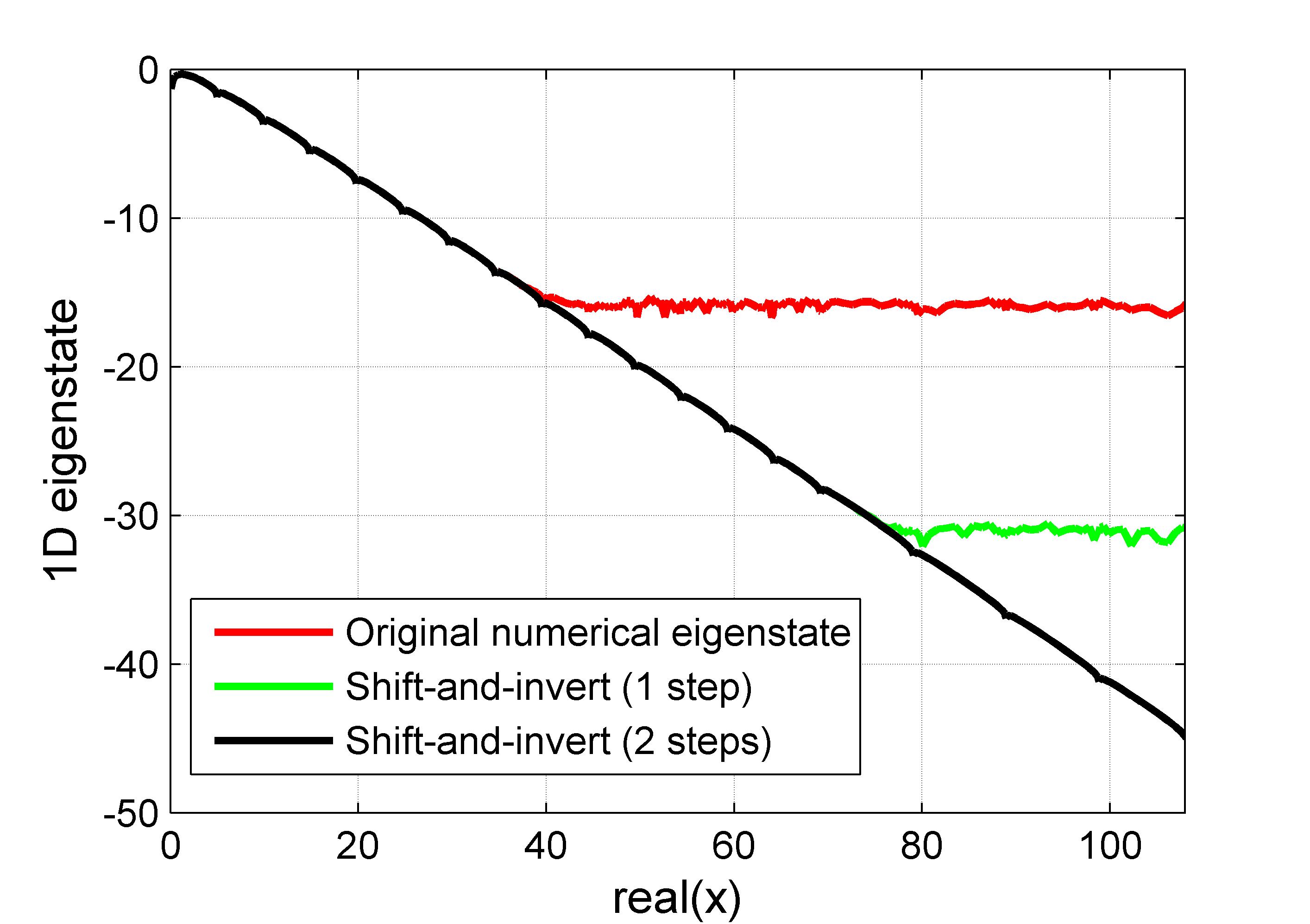}
  \caption{Modulus of the 1D eigenstate $\phi_n(x)$ calculated numerically along the complex contour with $\gamma = 9^{\circ}$. Plot in function of the real part of the complex-valued $x$-coordinate. Shown is the numerical eigenstate calculated `na\"ively' using a standard eigenfunction routine (red) and the eigenstate with improved accuracy using several steps of shift-and-invert (green and black). Round-off errors are eliminated after two applications of the shift-and-invert technique. Vertical axis in log-scale.}
  \label{fig:eig_mach_prec}
\end{center}
\end{figure}

\subsubsection{The 2D Temkin-Poet model: propagation of round-off errors in the eigenstates}

The Temkin-Poet (TP) model is used throughout this work as an
approximation to an impact ionization problem, where only $s$-waves are
used to describe the scattered electrons. The model has a long
history as a benchmark problem for Schr\"odinger
scattering problems, see e.g.\ \cite{rescigno1999collisional}, where a
two-dimensional TP model was used to describe the impact ionization of
molecular hydrogen.  Similarly, the impact ionization of helium can be
modelled by a 3D Temkin-Poet model \cite{vanroose2006double}.

%The TP model problem serves as the primary test case
%for numerical validation of the proposed MG-CCCS method in this paper.

We consider the 2D Temkin-Poet model discretized using a spectral
element type discretization; however, the following observations are
independent of the exact discretization scheme. The total wave
$u(x,y)$ is the sum of the incoming wave $u_{\text{in}}$ and the
scattered wave $u_{\text{sc}}$, i.e.  $u(x,y) = u_{\text{in}}(x,y) +
u_{\text{sc}}(x,y)$.  Hence, the far field integral \eqref{eq:inteq2}
can be effectively split up into a sum of two integrals: one over
$u_{\text{in}}$ and the other over $u_{\text{sc}}$, see \eqref{eq:inteq3}. 
Note that the incoming wave comprises part of the right-hand side in 
the scattered wave equation \eqref{eq:discrete_schrodinger}.

When computing the complex-valued scattered wave for impact-ionization models, 
some numerical subtleties require specific attention. Notably, the incoming wave is
\begin{equation}
u_{\text{in}}(x,y) = \phi_n(x) \sin(k_n y),
\end{equation}
with $k_n = \sqrt{2(E-\lambda_n)}$, where $\phi_n$ is an eigenstate of
the one-body operator \eqref{eq:1d_eigenstate} with a negative
eigenvalue $\lambda_n$.  On the real-valued domain the eigenstate
$\phi_n$ is an exponentially decaying function, since $\lambda_n < 0$
implies that $\phi_n$ asymptotically behaves as $\exp(\lambda_n
x)$. After complex coordinate rotation $\phi_n$ is still a decaying
function. However, when calculated numerically using a standard
eigenfunction routine, the exponential decay of $\phi_n$ is
unavoidably truncated at machine precision. This is illustrated in
Figure \ref{fig:eig_mach_prec}, where it is observed that from a
certain distance on the exponential decay stagnates at around
$1.0^{-16}$.

Subsequently the numerical representation of the eigenstate $\phi_n(x)$ 
is multiplied by $\sin(k_n y)$ to form the incoming wave $u_{\text{in}}$. 
However, on the complex rotated grid the function $\sin(k_n y)$ becomes an exponentially 
growing function in $y$, with the exponential growth being governed by the 
rotation angle $\gamma$. The truncated part of the eigenstate $\phi_n$ - that 
should be negligibly small - becomes substantial for larger values of $y$ 
through the multiplication. Hence, the numerical round-off of the 1D eigenstate 
$\phi_n$ produces an unwanted error which propagates through the algorithm.

The incoming wave $u_{\text{in}}$ is additionally multiplied by the potentials 
$V_1(x)$ and $V_{12}(x,y)$ to form the right-hand side of the scattered wave equation 
\eqref{eq:discrete_schrodinger}.
The resulting source function is heavily contaminated due to the numerical round-off 
errors in $\phi_n$, as is shown in the left panel of Figure \ref{fig:tot_mach_prec}.
The contaminations at the rightmost (outgoing) domain edges (where both $x \to \infty$ and $y \to \infty$) 
are due to the numerical round-off error phenomenon described above.

\subsubsection{Solution to the truncation errors: shift-and-invert}

\begin{figure}
\begin{center}
  \includegraphics[width=0.45\textwidth]{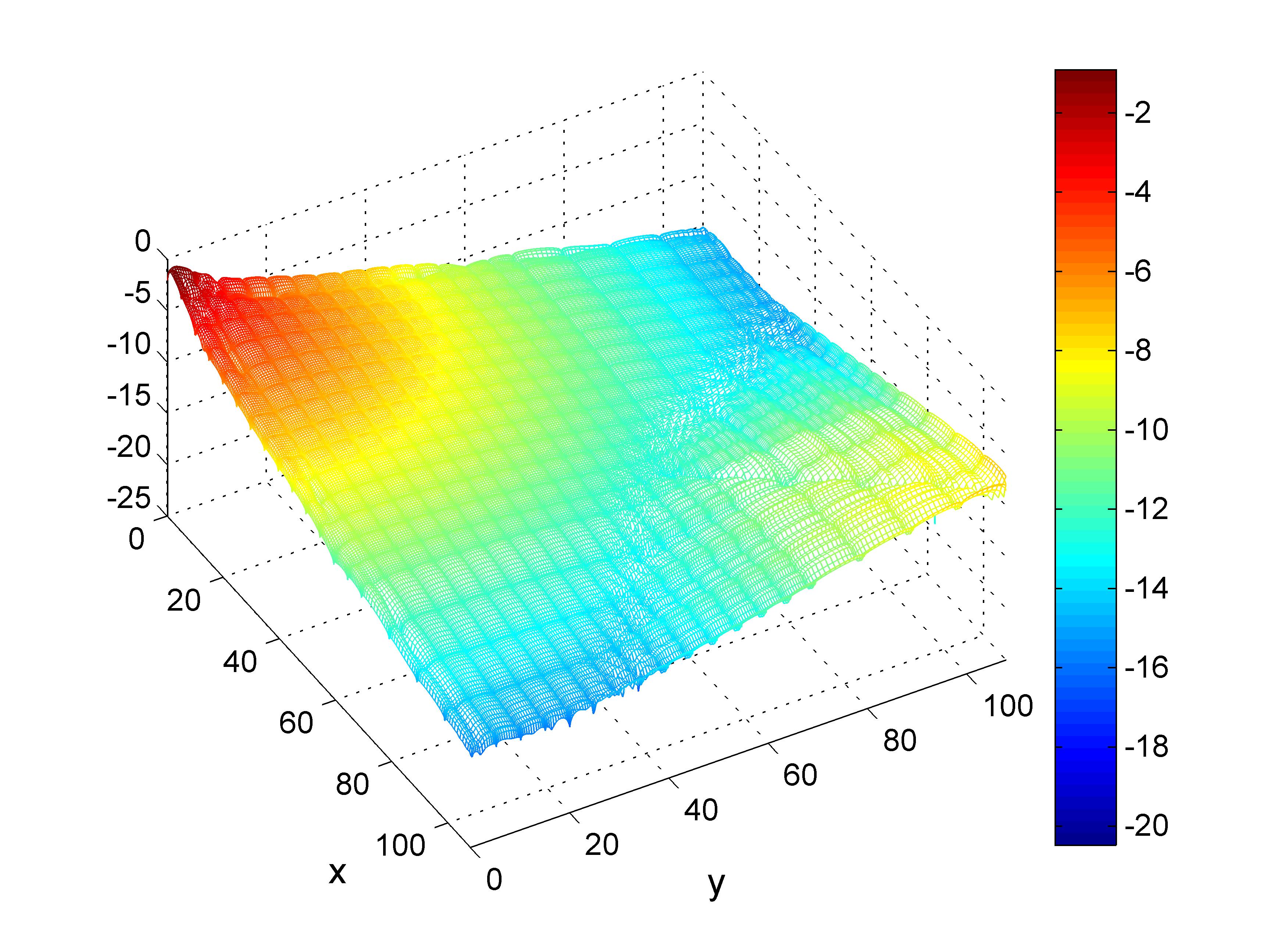}
  \includegraphics[width=0.45\textwidth]{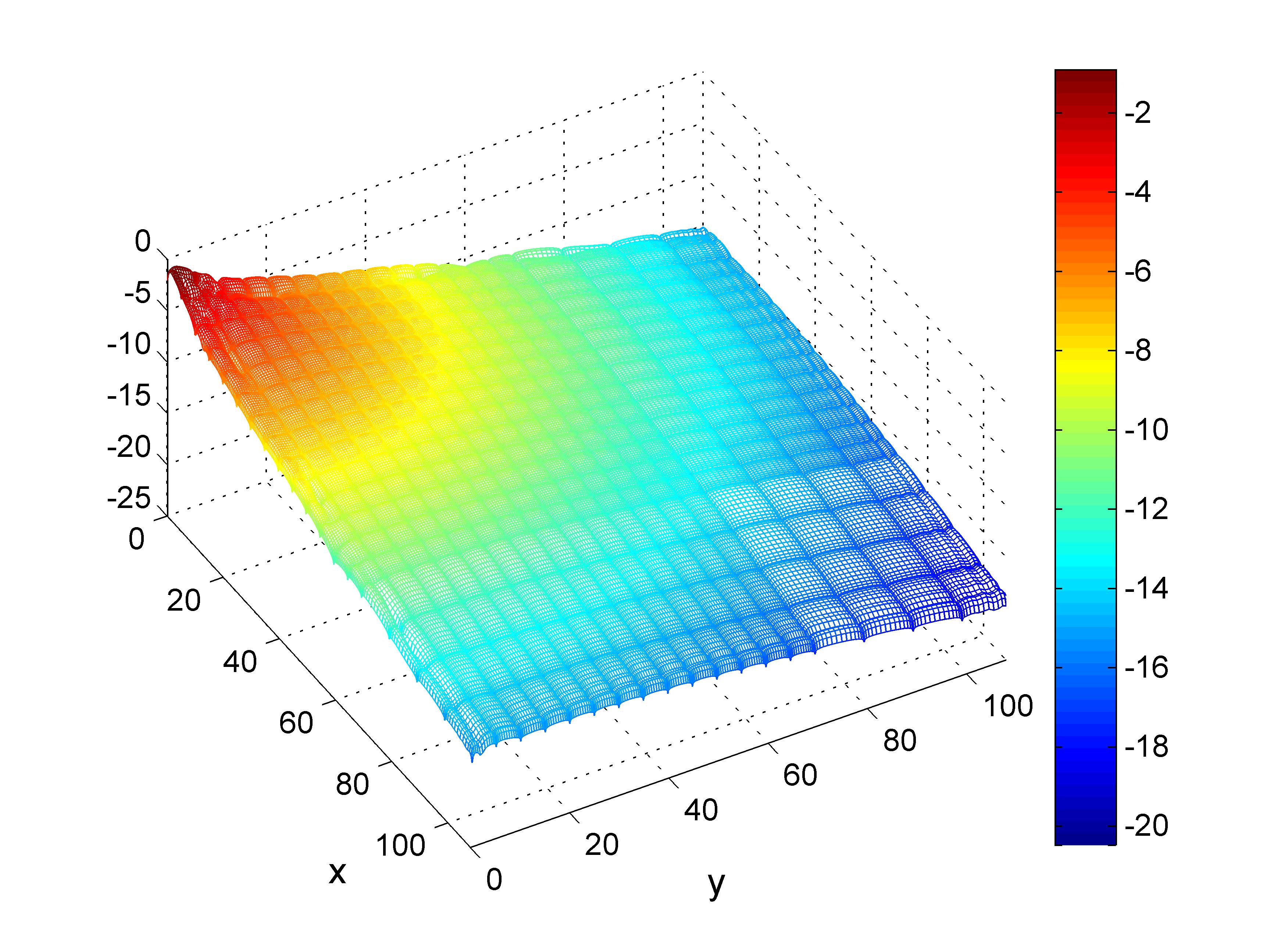}
  \caption{Modulus of the function $\left(V_1(x) - V_{12}(x,y)\right) \phi_n(x)
    \sin(k_n y)$, which forms the right-hand side of the scattered wave equation 
    \eqref{eq:discrete_schrodinger}, calculated numerically on the complex contour 
    with $\gamma = 9^{\circ}$ for energy $E = 1$. Potentials are $V_1(x) = -\exp(-x^2)$ 
    and $V_{12}(x,y) = \exp(-(x+y)^2)$; comparable results are obtained using Temkin-Poet 
    potentials. Plot in function of the real part of the complex-valued $x$- and $y$-coordinates. 
    Vertical axis in log-scale. Left: with standard eigenfunction computation of $\phi_n(x)$. 
    Round-off errors contaminate the function in the region  where $80<x$ and $80<y$, 
    cf.~Figure \ref{fig:eig_mach_prec}. Right: adapted computation using the iteratively 
    refined eigenstate $\phi_n(x)$. Round-off errors have been eliminated.}
  \label{fig:tot_mach_prec}
\end{center}
\end{figure}

To reduce the numerical inaccuracies due to the round-off of the 1D
eigenstates, we apply the classical concept of shift-and-invert
\cite{grimes1994shifted} to the eigenstate $\phi_n$, which is closely
related to iterative refinement \cite{wilkinson1994rounding}.  This
simple iterative update step (power iteration) is based on an
additional solve using the shifted Hamiltonians $(H_1 - \lambda_n)$ or
$(H_2 - \lambda_n)$, hence allowing for a more accurate representation
of the eigenstates $\phi_n$. The numerical eigenstate after
application of the shift-and-invert technique is shown on Figure
\ref{fig:eig_mach_prec}.  Using the improved accuracy of the
eigenstates, the round-off errors due to multiplication by the
exponentially increasing sine function can be avoided, as illustrated
in the right panel of Figure \ref{fig:tot_mach_prec}.  Note that the
complex rotation angle $\gamma$ plays an important role in the
numerical accuracy and the avoidance of round-off errors. Even  if the numerical accuracy of the eigenstates $\phi_n(x)$ and
$\varphi_n(y)$ is improved using the shift-and-invert technique as
suggested above, the multiplication by $\sin(k_n y)$ and $\sin(k_n x)$
respectively should be treated with care. Indeed, if $\gamma$ is
chosen too large, the exponential growth of these sine functions may
still introduce a blow-up of round-off errors in the computation. To
avoid this issue, the rotation angle $\gamma$ should always be chosen
as small as possible, under the condition that the multigrid scheme remains
stable. This trade-off in the choice of the complex rotation is
characteristic for complex shifted/rotated problems,
cf.\,\cite{erlangga2006novel,cools2013local}.

%---------------- SUBSECTION --------------------------------
\subsection{Validation of the complex contour approach: computation of the ionization cross sections}

\begin{figure}
\begin{center}
  \includegraphics[width=0.4\textwidth]{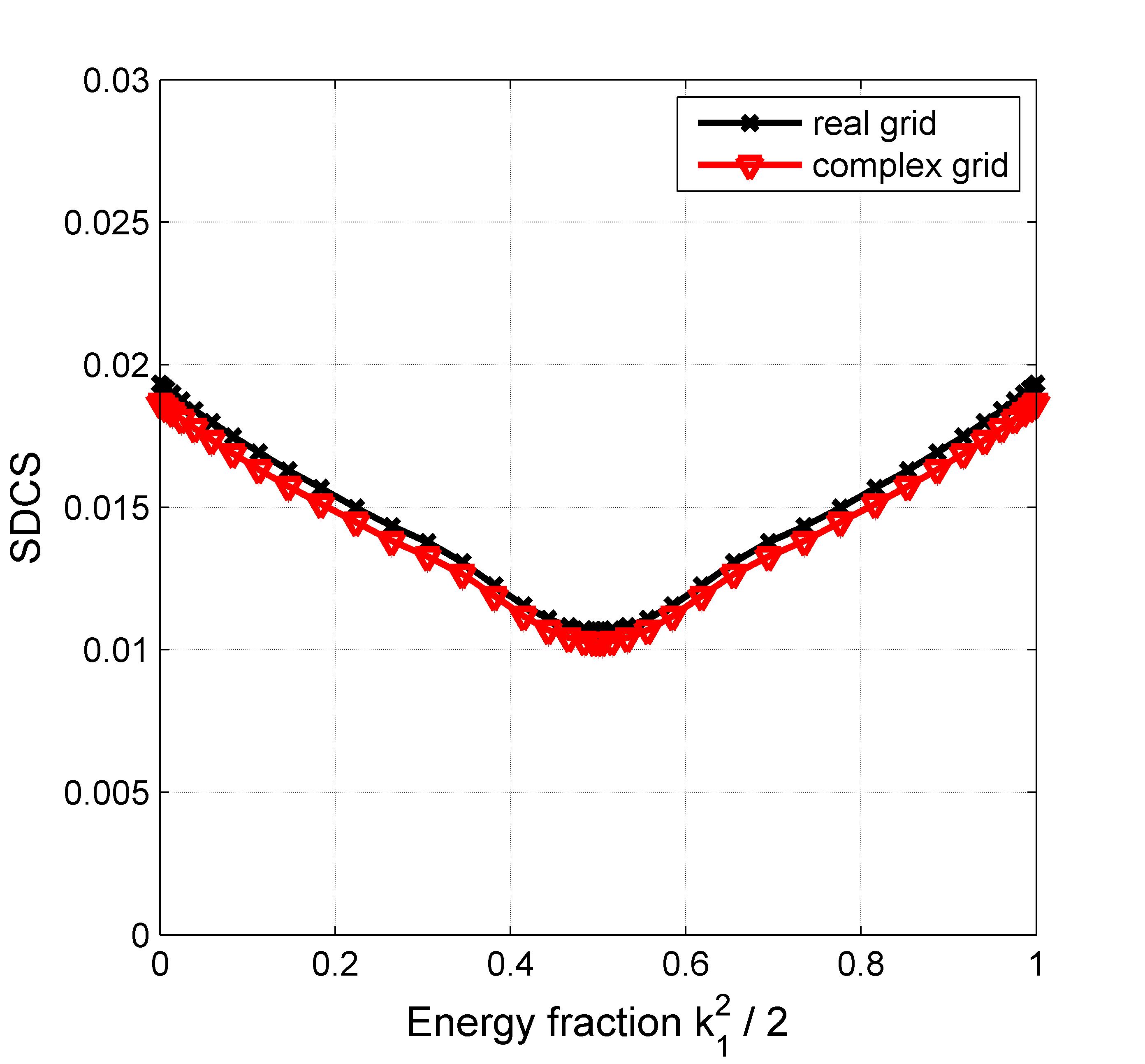} \qquad
  \includegraphics[width=0.4\textwidth]{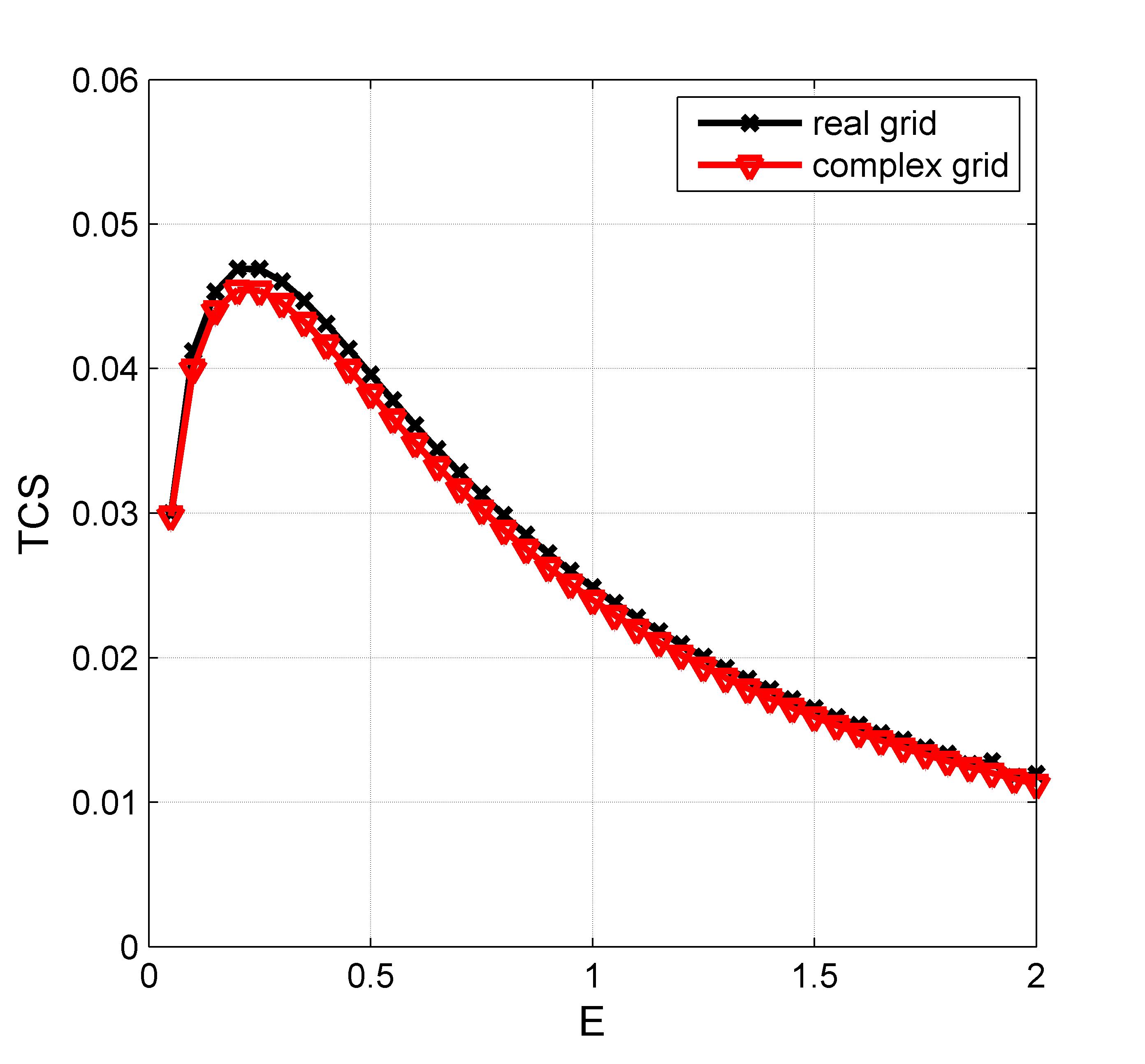}
  \caption{Left: Single differential cross section $\zeta(k_1,k_2)$ \eqref{eq:double-ionization} in function of the energy fraction $k_1^2/2 \in [0,1]$. 
  Right: Total double ionization cross section $\sigma_{tot}(E)$ \eqref{eq:total-double-ionization} in function of the energy $E \in [0,2]$. 
  Black: classical calculation on the real-valued domain using ECS boundaries with $\theta_{ECS}=30^{\circ}$. Red: alternative calculation on the complex-valued domain with $\gamma=9^{\circ}$.}
  \label{fig:CS}
\end{center}
\end{figure}

Following the discussion in the previous section, the cross section $\zeta(k_1,k_2)$ of the 2D Temkin-Poet model problem can now be calculated accurately. The TP model problem features potentials $V_1(x) = -1/x, \, V_2(y) = -1/y$ and $V_{12}(x,y) = 1/\max(x,y)$, and the right-hand side is given by $f(x,y) = xy\exp(-(x+y)^2)$. The double ionization cross section is given by expression \eqref{eq:double-ionization}, where $k_1 = \sqrt{2E}\sin(\alpha)$ and $k_2 = \sqrt{2E}\cos(\alpha)$. Hence, the cross section $\zeta(k_1,k_2)$ can be calculated for a single energy $E$ in function of the incoming wave angle $\alpha \in [0,\pi/2]$ to yield the single differential cross section (SDCS). The total cross section (TCS) $\sigma_{tot}(E)$ can then be calculated as an integral of $\zeta$ over all energy fractions and is given by expression \eqref{eq:total-double-ionization} in function of the energy of the system. 

Figure \ref{fig:CS} (left panel) shows the single differential cross section for the 2D Temkin-Poet model problem for the energy $E = 1$, at which a full break-up of the system (double ionization) occurs. The scattering solution is computed using a spectral element discretization of the domain $\Omega = [0,108]^2$ using $n_x = n_y = 269$ grid points in every spatial direction. The SDCS is computed on both the complex contour domain with $\gamma = 9^{\circ}$ and the classical real-valued domain with an ECS absorbing boundary layer ($\theta_{ECS} = 30^{\circ}$) for comparison. Note that to avoid propagation of numerical eigenstate round-off errors throughout the algorithm, the shift-and-invert technique is applied to the 1D eigenmodes to ensure numerical accuracy. The complex contour approach yields an accurate representation of the SDCS compared to the classical real-valued computation.

The total cross section for the 2D Temkin-Poet model problem is shown in Figure \ref{fig:CS} (right panel) for a range of energies $E \in [0,2]$ for which both single and double ionization occur simultaneously. The figure validates the complex contour approach by comparing the TCS computed using the classical real-valued domain to the computation of the cross section using the complex contour approach, where notably 2 steps of shift-and-invert are included for numerical accuracy. The resulting total cross section is close to identical for both methods.

%%%%%%%%%  SECTION  %%%%%%%%%%%%%%%%%%%%%%%%%%%%%%%%%%%%%%%%%%%%%%%%

\begin{figure}
\begin{center}
  \includegraphics[width=0.48\textwidth]{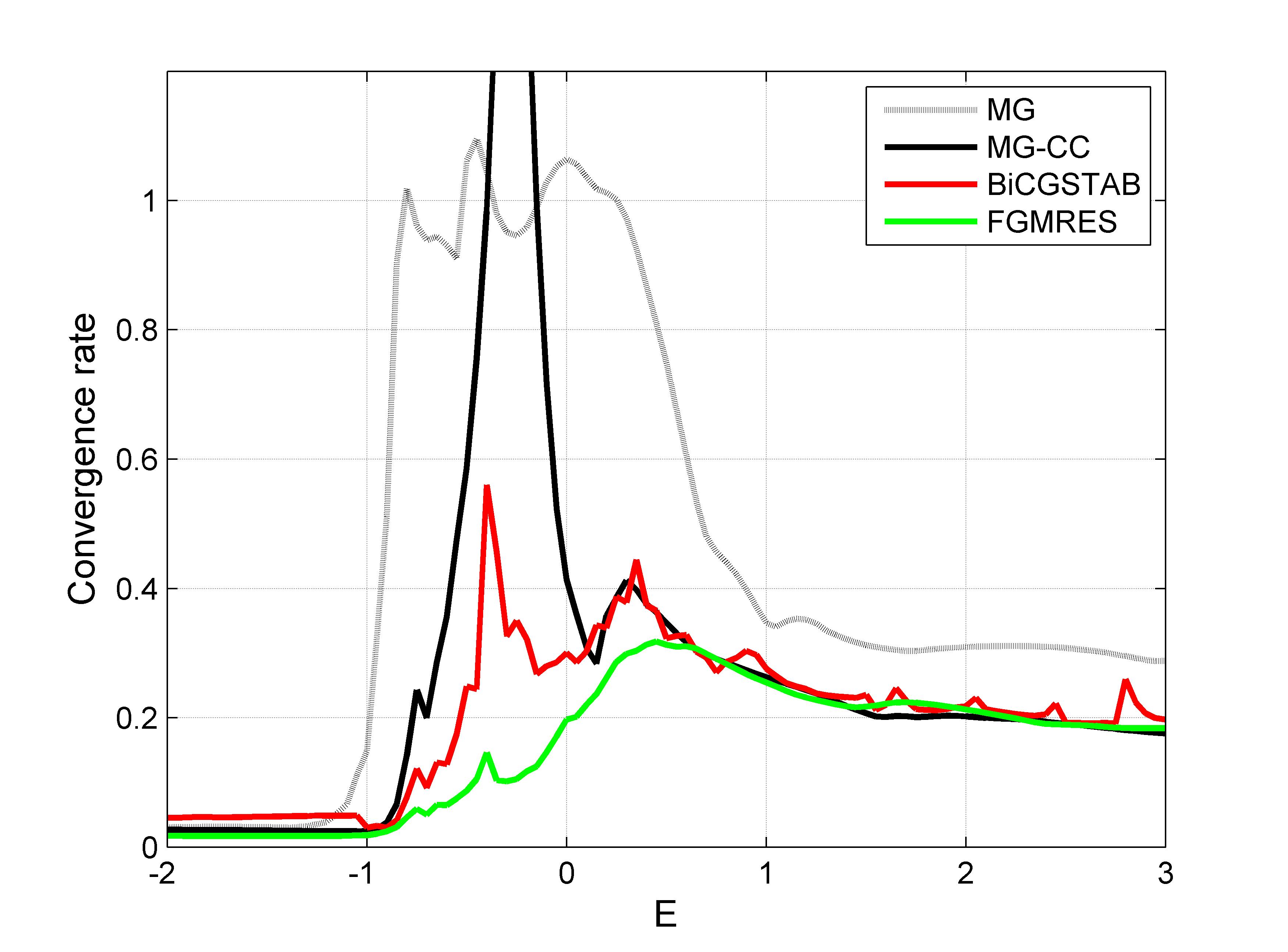}
  \includegraphics[width=0.48\textwidth]{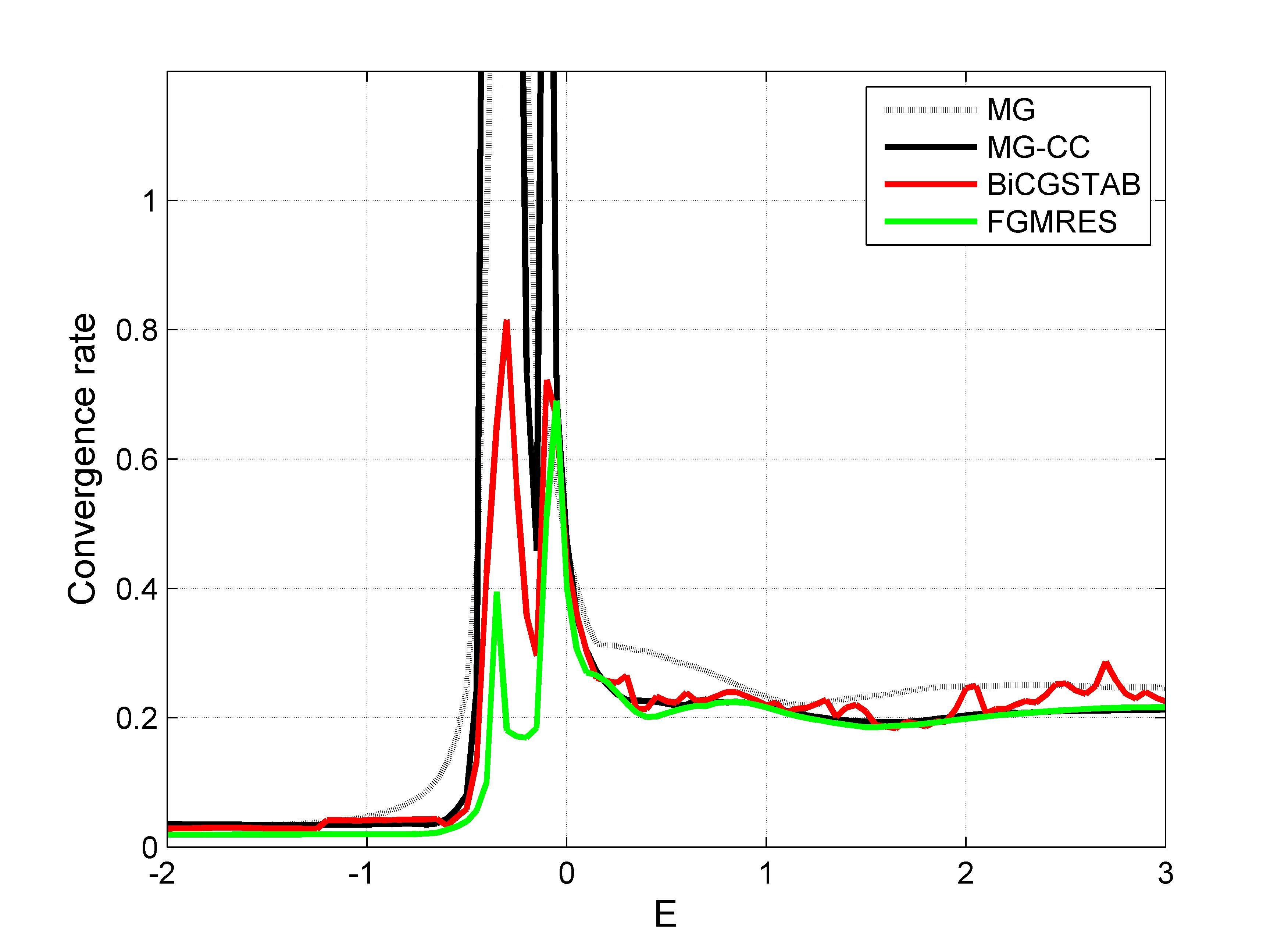}
  \caption{Comparison of the convergence rates of different iterative solution methods for two 2D complex contour Schr\"odinger model problems. Classical MG is used as a solver only. The MG-CCCS method (with $M = L = 2$) is used both as a stand-alone solver and Krylov preconditioner for BiCGSTAB and FGMRES. Left: exponential potentials model problem, domain $[0,20]^2$, no.\,grid points $n_x = n_y = 256$, no.\,MG levels $l = 8$. Right: Temkin-Poet model problem, domain $[0,100]^2$, no.\,grid points $n_x = n_y = 1024$, no.\,MG levels $l = 10$.}
  \label{fig:convrate2D}
\end{center}
\end{figure}

\section{Numerical results: convergence of the MG-CCCS method} \label{sec:numer}

In this section we validate the efficiency and scalability of the Multigrid-Coupled Channel (MG-CCCS) combined method for solving the damped Schr\"odinger equation along a complex contour. We illustrate the convergence of MG-CCCS both as a solver and a Krylov preconditioner. When used as a Krylov preconditioner, only one MG-CCCS V-cycle is used to approximately solve the preconditioning system, as is common practice in the multigrid literature \cite{erlangga2006novel,bollhoefer2009algebraic}. 
% Convergence results for two distinct model problems are presented: a 2D ionization problem with exponentially decaying potentials, and a 2D Temkin-Poet problem. 

%---------------- SUBSECTION --------------------------------
\subsection{The 2D ionization model with exponential potentials}

As a proof of concept, we first consider a simple 2D electron-impact ionization model problem with exponentially decreasing potentials. The potentials are given by $V_1(x) = -4.5\exp(-x^2)$, 
$V_2(y) = -4.5\exp(-y^2)$ and $V_{12}(x,y) = 2\exp(-0.1(x+y)^2)$. The right-hand side is $f(x,y) = \exp(-3(x+y)^2)$. The corresponding driven Schr\"odinger equation \eqref{eq:partialwave2d} is solved along a complex contour with rotation angle $\gamma = \pi/18 = 10^{\circ}$ on a uniform discretization of the domain $\Omega = [0,20]^2$ featuring $n_x = n_y = 256$ grid points in every spatial direction. This ensures a minimum of 20 grid points per wavelength for all energies $E \in [-2,3]$, yielding an accurate representation of the waveforms.

Figure \ref{fig:convrate2D} (left panel) compares the convergence rate of the classical multigrid method to the MG-CCCS, MG-CCCS BiCGSTAB and FGMRES solvers in function of the energy. The MG scheme denotes the standard multigrid method using V(1,1)-cycles with a GMRES(3) pre- and post-smoothing substitute. The MG-CCCS scheme includes a Coupled Channel correction step with $L = M = 2$ after each V-cycle. The classical MG solver is instable in the region of energies for which only single ionization occurs, i.e.\,for $E \in [-1,0]$, cf.~conclusions drawn in \cite{cools2014efficient}. In principle the energy regimes for which only single ionization occurs do not require a full 2D description, since a 1D model would suffice to describe the single ionization waveforms. However, rather than adapting the mathematical model description to the energy $E$, we in this work aim at constructing a robust 2D solver that is effective for \emph{all} energy regimes. 

The MG-CCCS solver features improved convergence for most energy levels due to the additional elimination of 1D evanescent modes from the error. This implies the overall convergence of the MG-CCCS scheme is significantly better than the standard MG convergence rate. Stability of the MG-CCCS scheme is unfortunately still not guaranteed for every energy. However, the MG-CCCS scheme can be used as a preconditioner to a general Krylov method to further improve robustness. Acceleration of the MG-CCCS scheme by a governing Krylov solver leads to a working solution method over the entire energy range. The MG-CCCS preconditioned BiCGSTAB solver features a convergence rate of less than $0.6$ for all energy regimes, while the restarted FGMRES(5) features a convergence rate of at most $0.32$ for all energies. The MG-CCCS Krylov solver hence proves to be a robust solution method for the 2D ionization model problem.

%---------------- SUBSECTION --------------------------------

\subsection{The 2D Temkin-Poet model}

\begin{figure} 
\begin{center}
  \includegraphics[width=0.7\textwidth]{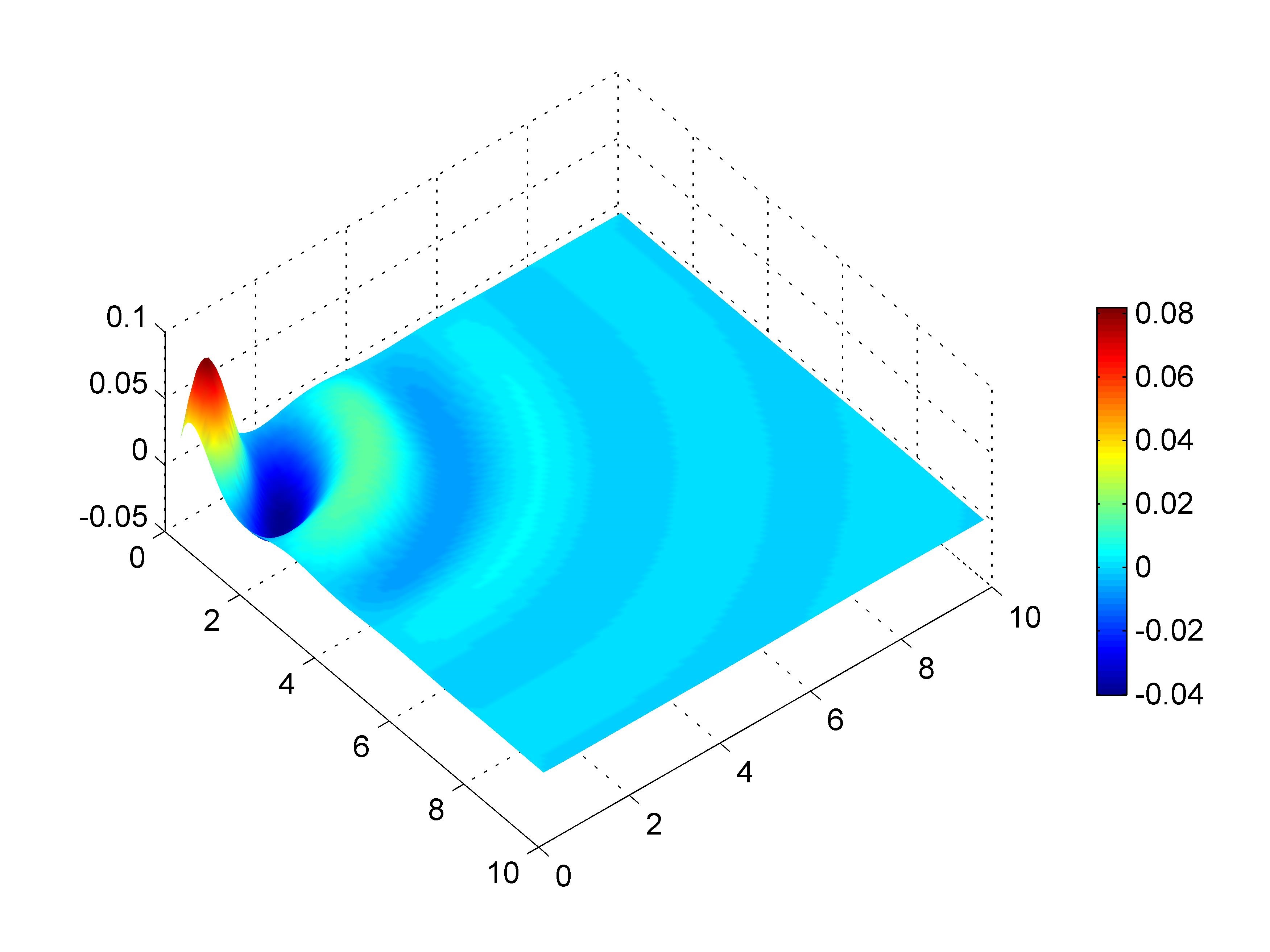} \vspace{-0.2cm}
  \caption{Numerical solution $u^N$ to the 2D complex-valued Temkin-Poet model problem with energy $E = 1$. 
  Discretization of the domain $\Omega = [0,200]^2$ using a total of $2048^2$ grid points.
  Displayed is a close-up of the solution on the subdomain $[0,10]^2$. The exponential decay of the solution on the complex rotated grid is clearly visible. 
  Solution method details: see Table \ref{tab:TP}.}
  \label{fig:sol_cont}
\end{center}
\end{figure}

\begin{table}
\centering
\begin{tabular}{| c | c | c  c  c  c  c |}
\cline{2-7}
	\multicolumn{1}{c|}{}  & domain $\Omega$ \rule{0pt}{2.6ex}	& $[0,10]^2$ & $[0,20]^2$ & $[0,50]^2$ & $[0,100]^2$ & $[0,200]^2$ \\
	\multicolumn{1}{c|}{} & $n_x \times n_y$ 								  & $128^2$ & $256^2$ & $512^2$ & $1024^2$ & $2048^2$ \\
\hline
	\multirow{3}{*}{$E = -2$} & iterations& 3 & 3 & 3 & 3 & 3 \\
  													& total CPU time	& 0.3 s. & 0.9 s. & 3.8 s. & 15.1 s. & 59.4 s.  \\
  													& time/gridpoint	& 18.3 $\mu$s. & 13.7 $\mu$s. & 14.5 $\mu$s. & 14.4 $\mu$s. & 14.2 $\mu$s.  \\
\hline													
	\multirow{3}{*}{$E = -1$} & iterations& 3 & 3 & 3 & 3 & 3 \\
  													& total CPU time	& 0.3 s. & 0.9 s. & 3.9 s. & 15.2 s. & 60.3 s.  \\
  													& time/gridpoint	& 18.3 $\mu$s. & 13.7 $\mu$s. & 14.9 $\mu$s. & 14.5 $\mu$s. & 14.4 $\mu$s.  \\
\hline
	\multirow{3}{*}{$E = 0$} 	& iterations& 5 & 9 & 10 & 11 & 11 \\
  													& total CPU time	& 0.5 s. & 2.5 s. & 13.0 s. & 56.2 s. & 227.1 s.  \\
														& time/gridpoint	& 30.5 $\mu$s. & 38.1 $\mu$s. & 49.6 $\mu$s. & 53.6 $\mu$s. & 54.1 $\mu$s.  \\
\hline
	\multirow{3}{*}{$E = 1$} 	& iterations& 6 & 7 & 7 & 7 & 7 \\
  													& total CPU time	& 0.5 s. & 2.1 s. & 9.0 s. & 35.8 s. & 145.1 s.  \\
  													& time/gridpoint	& 30.5 $\mu$s. & 32.0 $\mu$s. & 34.3 $\mu$s. & 34.1 $\mu$s. & 34.6 $\mu$s.  \\
\hline
	\multirow{3}{*}{$E = 2$} 	& iterations& 7 & 7 & 7 & 7 & 7 \\
  													& total CPU time	& 0.6 s. & 2.0 s. & 9.0 s. & 35.4 s. & 140.4 s.  \\
														& time/gridpoint	& 36.6 $\mu$s. & 30.5 $\mu$s. & 34.3 $\mu$s. & 33.8 $\mu$s. & 33.5 $\mu$s.  \\
\hline
	\multirow{3}{*}{$E = 3$}  & iterations& 7 & 7 & 8 & 8 & 8 \\
  													& total CPU time	& 0.6 s. & 2.0 s. & 10.3 s. & 40.1 s. & 162.0 s.  \\
  													& time/gridpoint	& 36.6 $\mu$s. & 30.5 $\mu$s. & 39.3 $\mu$s. & 38.2 $\mu$s. & 38.6 $\mu$s.  \\
\hline
\end{tabular}
\caption{Scalability results for the 2D complex-valued Temkin-Poet model problem for energies $E = -2,-1,0,1,2$ and $3$. MG-CCCS preconditioned FGMRES ($M = L = 2$) iteration count, total CPU time and average CPU time per gridpoint required to solve the problem up to a relative residual tolerance of $10^{-6}$ in function of the problem size. Discretizations respecting the 20 points per wavelength criterion.}
\label{tab:TP}
\end{table}

We now consider a more realistic Temkin-Poet benchmark problem on a large domain $\Omega = [0,100]$, requiring a fine discretization of $n_x = n_y = 1024$ grid points to satisfy the wavenumber criterion of 20 points per wavelength for all energies $E \in [-2,3]$. The Temkin-Poet potentials are $V_1(x) = -1/x, \, V_2(y) = -1/y$ and $V_{12}(x,y) = 1/\max(x,y)$, and the right-hand side for this problem is again $f(x,y) = \exp(-3(x+y)^2)$.

The convergence rates for the 2D Temkin-Poet model are shown in the right panel of Figure \ref{fig:convrate2D}. Note how the effect of the Coupled Channel correction is somewhat less pronounced for the TP model problem. Indeed, a smaller improvement in convergence rate is measured when comparing the classical MG and MG-CCCS schemes. However, the MG-CCCS preconditioned Krylov solvers again provide a stable solution method for all energies. Convergence rates for the MG-CCCS FGMRES solver generally lie below $0.30$, with the exception of a small outlier at slightly negative energies where it rises to around $0.70$. Convergence of the MG-CCCS Krylov solver is guaranteed for all $E \in [-2,3]$, yielding a robust solver for all energy regimes.
 
Table \ref{tab:TP} shows the scalability of the MG-CCCS preconditioned FGMRES solver for the Temkin-Poet model problem for different fixed energy levels $E = -2, -1, 0, 1, 2$ and $3$. Note that the energies $E = -2, -1, 0$ correspond to single ionization regime, whereas the energies $E = 1, 2, 3$ represent regions of physical interest where single and double ionization simultaneously occur. The number of MG-CCCS FGMRES iterations, the total CPU time, and the average time per grid point\footnote{System specifications: Intel Core i7-2720QM 2.20GHz CPU, 6MB Cache, 8GB RAM.} required to solve the problem up to a relative residual tolerance of $10^{-6}$ are displayed in function of the number of grid points (increasingly more accurate discretization). Perfect $\mathcal{O}(N)$ scalability can be observed from the table for all energies, except for $E = 0$, where a small convergence deterioration is observed. The number of Krylov iterations remains constant as the number of discretization points increases, resulting in a CPU time which effectively scales linearly in the number of unknowns. Moreover, good scalability in function of the energy level is observed. Hence, it is concluded that the MG-CCCS preconditioned Krylov method is not only robust to different energies as shown by Figure \ref{fig:convrate2D}, it additionally exhibits optimal scaling properties in function of the problem size.

%%%%%%%%%  SECTION  %%%%%%%%%%%%%%%%%%%%%%%%%%%%%%%%%%%%%%%%%%%%%%%%

\section{Conclusions} \label{sec:concl}

In this paper we have developed an efficient and robust
multigrid-based computational scheme for the calculation of the
ionization cross sections of a Schr\"odinger-type quantum mechanical
break-up problem. The proposed method achieves linear scalability 
in the number of unknowns. It is based on the so-called complex contour
approach \cite{cools2014efficient}, i.e.~a reformulation of the
classical real-valued far field integral (cross section) to an
integral over a complex rotated domain. The numerical solution to the
underlying driven Schr\"odinger system (scattered wave) on the
complex-valued domain is much easier to obtain iteratively, which
overcomes the primary computational bottleneck for the calculation of
the ionization cross sections. 

The MG-CCCS method proposed in this work combines the functionality 
of multigrid on the complex-rotated (damped) Schr\"odinger problem with 
a Coupled Channel Correction Scheme \cite{heller1973comment,mccarthy1983momentum}, 
which accounts for the existence of single ionization waves in the solution 
that might deteriorate the multigrid convergence. Potentially accelerated 
by a Krylov subspace method for improved robustness, the MG-CCCS solver is 
capable of efficient computation of the complex-valued scattered wave for 
any energy $E$, featuring optimal $\mathcal{O}(N)$ scalability in the number 
of unknowns.  Furthermore, the Coupled Channel Correction Scheme improves the 
stability of the classical multigrid method, ensuring robustness of the MG-CCCS 
Krylov solver with respect to the energy of the system. 

Note that there are other possibly remedies for removing the remaining errors
caused by the presence of single ionization waves along the axes. For
example, a semi-coarsening strategy that removes grid points in the
middle of the domain but keeps sufficient points along the axes might
also efficiently remove these errors. Furthermore, note that the 
Coupled Channel Correction Scheme proposed in this work shows resemblance 
to a deflation strategy. In this paper the CCCS was implemented as an explicit additional 
step to increase the performance of the multigrid preconditioner.
However, it could alternatively be embedded in the governing Krylov solver 
in the form of a deflation step, cf.~the work by Sheikh et al.~\cite{sheikh2013convergence}, 
which analogously aims at eliminating problematic 
eigenstates in the solution and may lead to an even more efficient solver.

Particular care is advised when solving impact ionization problems on 
the complex-valued domain, since the complex integral implementation is prone 
to round-off errors due to the exponential decay and growth of the eigenstates 
and incoming wave components respectively. Consequently, numerical round-off 
errors might propagate through the algorithm and contaminate the resulting 
cross sections when not treated properly. We suggested the use of a shift-and-invert 
technique to ensure proper accuracy of the eigenstate functions for impact ionization simulations. 

The choice of the complex rotation angle is of fundamental importance in this regard, 
since it governs the rate of exponential growth of particular integrand components. 
Although from the analytical viewpoint there are no fundamental restrictions 
on the magnitude of the complex rotation, and despite the fact that the multigrid solver clearly benefits 
from a heavy damping of the problem, we have shown that choosing the rotation angle 
too large may result in numerical instabilities when calculating the far field integral. 
Hence, the rotation angle is bounded both 
from below by the requirement of a stable multigrid solver, and from above by the 
numerical accuracy of the implementation. A mindful trade-off between these two conditions 
should be achieved in order to obtain a functional computational scheme.

As shown in this paper, the MG-CCCS method allows for a linear scaling of the computational 
time in function of the number of unknowns. Sublinear scaling may be achieved by a
parallel implementation of the multigrid solver, which will be treated in future work.

Numerical results on the 2D Temkin-Poet benchmark problem validate the
proposed MG-CCCS complex contour approach, showing the method to yield
an accurate representation of the single differential and total double
ionization cross sections for the two-body electron break-up
problem. Although the results provided in this paper are restricted to
2D model problems for notational and implementation convenience, the
extension of the proposed MG-CCCS method to higher spatial dimensions,
notably the three-particle break-up system, is straightforward and
will also be treated as part of future work.

\section*{Acknowledgments} \label{sec:ackno}
\noindent This research is funded by the Research Council of the
University of Antwerp. The authors additionally thank Bram Reps for
fruitful discussions on the subject.

\nocite{*}
{\footnotesize
\bibliographystyle{plain}
\bibliography{refs}
}

\end{document}

%% file: total.tex
\begin{tikzpicture}
\begin{axis}[
        xlabel=$E$,
        ylabel=Cross section,
        ymax=0.0045]

                \addplot[smooth,black,dashed] plot coordinates{
(-1.0215007e+00   ,  0 		   )
(-9.2150069e-01   ,  6.3858252e-04)
(-8.2150069e-01   ,  9.2467986e-04)
(-7.2150069e-01   ,  1.0749516e-03)
(-6.2150069e-01   ,  1.1890037e-03)
(-5.2150069e-01   ,  1.2934345e-03)
(-4.2150069e-01   ,  1.3972521e-03)
(-3.2150069e-01   ,  1.4765268e-03)
(-2.2150069e-01   ,  1.5438088e-03)
(-1.2150069e-01   ,  1.6045563e-03)
(-2.1500689e-02   ,  1.6558229e-03)
( 7.8499311e-02   ,  1.6941891e-03)
( 1.7849931e-01   ,  1.7290531e-03)
( 2.7849931e-01   ,  1.7594438e-03)
( 3.7849931e-01   ,  1.7788039e-03)
( 4.7849931e-01   ,  1.7946639e-03)
( 5.7849931e-01   ,  1.8132298e-03)
( 6.7849931e-01   ,  1.8252779e-03)
( 7.7849931e-01   ,  1.8288557e-03)
( 8.7849931e-01   ,  1.8339955e-03)
( 9.7849931e-01   ,  1.8431227e-03)
( 1.0784993e+00   ,  1.8474371e-03)
( 1.1784993e+00   ,  1.8438096e-03)
( 1.2784993e+00   ,  1.8406952e-03)
( 1.3784993e+00   ,  1.8437351e-03)
( 1.4784993e+00   ,  1.8469109e-03)
( 1.5784993e+00   ,  1.8426860e-03)
( 1.6784993e+00   ,  1.8337866e-03)
( 1.7784993e+00   ,  1.8289090e-03)
( 1.8784993e+00   ,  1.8302016e-03)
( 1.9784993e+00   ,  1.8307270e-03)
( 2.0784993e+00   ,  1.8241433e-03)
( 2.1784993e+00   ,  1.8130237e-03)
( 2.2784993e+00   ,  1.8052771e-03)
( 2.3784993e+00   ,  1.8042338e-03)
( 2.4784993e+00   ,  1.8049649e-03)
( 2.5784993e+00   ,  1.8003346e-03)
( 2.6784993e+00   ,  1.7892010e-03)
( 2.7784993e+00   ,  1.7774952e-03)
( 2.8784993e+00   ,  1.7714823e-03)
( 2.9784993e+00   ,  1.7711985e-03)};
\addlegendentry{Single ionization (real)}  

\addplot[mark=triangle*,black,only marks] plot coordinates{
(-1.0215007e+00  ,            0 )
(-8.2150069e-01  ,  9.0468611e-04 )
(-6.2150069e-01  ,  1.2154105e-03 )
(-4.2150069e-01  ,  1.4230472e-03 )
(-2.2150069e-01  ,  1.5720915e-03 )
(-2.1500689e-02  ,  1.6830472e-03 )
( 1.7849931e-01  ,  1.7563615e-03 )
( 3.7849931e-01  ,  1.8025587e-03 )
( 5.7849931e-01  ,  1.8385737e-03 )
( 7.7849931e-01  ,  1.8504130e-03 )
( 9.7849931e-01  ,  1.8668001e-03 )
( 1.1784993e+00  ,  1.8628721e-03 )
( 1.3784993e+00  ,  1.8659478e-03 )
( 1.5784993e+00  ,  1.8603197e-03 )
( 1.7784993e+00  ,  1.8479752e-03 )
( 1.9784993e+00  ,  1.8494916e-03 )
( 2.1784993e+00  ,  1.8276939e-03 )
( 2.3784993e+00  ,  1.8228369e-03 )
( 2.5784993e+00  ,  1.8154194e-03 )
( 2.7784993e+00  ,  1.7906321e-03 )
( 2.9784993e+00  ,  1.7882720e-03 )
};
\addlegendentry{Single ionization (complex)}

\addplot[smooth,black,dotted] plot coordinates{
(-1.0215007e+00 ,  0 	          ) 
(-9.2150069e-01 ,  0            )
(-8.2150069e-01 ,  0		  )
(-7.2150069e-01 ,  0		  )
(-6.2150069e-01 ,  0 	          )
(-5.2150069e-01 ,  0 	          )
(-4.2150069e-01 ,  0 	          )
(-3.2150069e-01 ,  0		  )
(-2.2150069e-01 ,  0		  )
(-1.2150069e-01 ,  0    	  )
(-2.1500689e-02 ,  0		  )
( 7.8499311e-02 ,  1.1050837e-06)
( 1.7849931e-01 ,  5.3082119e-06)
( 2.7849931e-01 ,  1.2525175e-05)
( 3.7849931e-01 ,  2.2189419e-05)
( 4.7849931e-01 ,  3.4830544e-05)
( 5.7849931e-01 ,  4.9538427e-05)
( 6.7849931e-01 ,  6.5792520e-05)
( 7.7849931e-01 ,  8.3338714e-05)
( 8.7849931e-01 ,  1.0254151e-04)
( 9.7849931e-01 ,  1.2329677e-04)
( 1.0784993e+00 ,  1.4476612e-04)
( 1.1784993e+00 ,  1.6650046e-04)
( 1.2784993e+00 ,  1.8876351e-04)
( 1.3784993e+00 ,  2.1186457e-04)
( 1.4784993e+00 ,  2.3579776e-04)
( 1.5784993e+00 ,  2.6056731e-04)
( 1.6784993e+00 ,  2.8630097e-04)
( 1.7784993e+00 ,  3.1297130e-04)
( 1.8784993e+00 ,  3.4040841e-04)
( 1.9784993e+00 ,  3.6840341e-04)
( 2.0784993e+00 ,  3.9650813e-04)
( 2.1784993e+00 ,  4.2401240e-04)
( 2.2784993e+00 ,  4.5048901e-04)
( 2.3784993e+00 ,  4.7617820e-04)
( 2.4784993e+00 ,  5.0158394e-04)
( 2.5784993e+00 ,  5.2680940e-04)
( 2.6784993e+00 ,  5.5148539e-04)
( 2.7784993e+00 ,  5.7534142e-04)
( 2.8784993e+00 ,  5.9867209e-04)
( 2.9784993e+00 ,  6.2216583e-04)
};
\addlegendentry{Double ionization (real)}

\addplot[mark=*,color=black,only marks] plot coordinates{
( -1.0215007e+00  ,  0  )
( -8.2150069e-01  ,  0  )
( -6.2150069e-01  ,  0  )
( -4.2150069e-01  ,  0  )
( -2.2150069e-01  ,  0  )
( -2.1500689e-02  ,  0  )
(  1.7849931e-01  ,   5.2937397e-06  )
(  3.7849931e-01  ,   2.2421628e-05  )
(  5.7849931e-01  ,   4.8932351e-05  )
(  7.7849931e-01  ,   8.2720542e-05  )
(  9.7849931e-01  ,   1.2196342e-04  )
(  1.1784993e+00  ,   1.6624666e-04  )
(  1.3784993e+00  ,   2.1447172e-04  )
(  1.5784993e+00  ,   2.6423830e-04  )
(  1.7784993e+00  ,   3.1460554e-04  )
(  1.9784993e+00  ,   3.6522422e-04  )
(  2.1784993e+00  ,   4.1697918e-04  )
(  2.3784993e+00  ,   4.6947186e-04  )
(  2.5784993e+00  ,   5.2275418e-04  )
(  2.7784993e+00  ,   5.7544498e-04  )
(  2.9784993e+00  ,   6.2731293e-04  )
}
;
\addlegendentry{Double ionization (complex)}

        \addplot[smooth,black] plot coordinates{
(  -1.0215007e+00,   1.1366667e-05) 
(  -9.2150069e-01,   6.9686220e-04) 
(  -8.2150069e-01,   9.0293558e-04) 
(  -7.2150069e-01,   1.0658949e-03) 
(  -6.2150069e-01,   1.1920408e-03) 
(  -5.2150069e-01,   1.2972055e-03) 
(  -4.2150069e-01,   1.3965877e-03) 
(  -3.2150069e-01,   1.4774532e-03) 
(  -2.2150069e-01,   1.5455682e-03) 
(  -1.2150069e-01,   1.6060515e-03) 
(  -2.1500689e-02,   1.6569185e-03) 
(   7.8499311e-02,   1.6981357e-03) 
(   1.7849931e-01,   1.7370135e-03) 
(   2.7849931e-01,   1.7737332e-03) 
(   3.7849931e-01,   1.8051452e-03) 
(   4.7849931e-01,   1.8346068e-03) 
(   5.7849931e-01,   1.8658103e-03) 
(   6.7849931e-01,   1.8946983e-03) 
(   7.7849931e-01,   1.9193756e-03) 
(   8.7849931e-01,   1.9444192e-03) 
(   9.7849931e-01,   1.9721788e-03) 
(   1.0784993e+00,   1.9987841e-03) 
(   1.1784993e+00,   2.0213599e-03) 
(   1.2784993e+00,   2.0432017e-03) 
(   1.3784993e+00,   2.0682682e-03) 
(   1.4784993e+00,   2.0948876e-03) 
(   1.5784993e+00,   2.1185445e-03) 
(   1.6784993e+00,   2.1387604e-03) 
(   1.7784993e+00,   2.1596717e-03) 
(   1.8784993e+00,   2.1841138e-03) 
(   1.9784993e+00,   2.2098808e-03) 
(   2.0784993e+00,   2.2327429e-03) 
(   2.1784993e+00,   2.2518907e-03) 
(   2.2784993e+00,   2.2708371e-03) 
(   2.3784993e+00,   2.2930331e-03) 
(   2.4784993e+00,   2.3178593e-03) 
(   2.5784993e+00,   2.3414250e-03) 
(   2.6784993e+00,   2.3609632e-03) 
(   2.7784993e+00,   2.3778385e-03) 
(   2.8784993e+00,   2.3959541e-03) 
(   2.9784993e+00,   2.4176249e-03) 
};
\addlegendentry{Total cross section}

\end{axis}
\end{tikzpicture}